\documentclass[aps,prl,twocolumn,superscriptaddress,floatfix,longbibliography,showpacs]{revtex4-1}

\usepackage{graphicx}
\usepackage{dcolumn}
\usepackage{bm}
\usepackage{amssymb}
\usepackage{microtype}
\usepackage{gensymb}
\usepackage{xcolor}
\usepackage{float}
\usepackage{xfrac}
\usepackage{amsmath}
\usepackage{soul}

 \usepackage[columnwise]{lineno}
 \usepackage{lineno}

\newcommand{\beginsupplement}{
        \setcounter{table}{0}
        \renewcommand{\thetable}{S\arabic{table}}%
        \setcounter{figure}{0}
        \renewcommand{\thefigure}{S\arabic{figure}}%
     }

\newcommand{\TiCu}{Ti\textsubscript{3}Cu\textsubscript{4}}

\begin{document}


\title{Field-induced quantum critical point in the new itinerant antiferromagnet Ti$_3$Cu$_4$}

\author{Jaime M. Moya}
\affiliation{Applied Physics Graduate Program, Rice University, Houston, TX, 77005 USA}
\affiliation{Department of Physics and Astronomy, Rice University, Houston, TX, 77005 USA}

\author{Alannah M. Hallas}
\affiliation{Department of Physics and Astronomy, Rice University, Houston, TX, 77005 USA}
\affiliation{Department of Physics and Astronomy and Quantum Matter Institute, University of British Columbia, Vancouver, British Columbia V6T 1Z1, Canada}

\author{Vaideesh Loganathan}
\affiliation{Department of Physics and Astronomy, Rice University, Houston, TX, 77005 USA}

\author{C.-L. Huang}
\affiliation{Department of Physics and Astronomy, Rice University, Houston, TX, 77005 USA}
\affiliation{Department of Physics and Center for Quantum Frontiers of Research \& Technology (QFort), National Cheng Kung University, Tainan 701, Taiwan}

\author{Lazar L. Kish} 
\affiliation{Department of Physics and Materials Research Laboratory, University of Illinois at Urbana-Champaign, Urbana, Illinois 61801, USA}

\author{Adam A. Aczel} 
\affiliation{Neutron Scattering Division, Oak Ridge National Laboratory, Oak Ridge, Tennessee 37831, USA}

\author{J. Beare}
\affiliation{Department of Physics and Astronomy, McMaster University, Hamilton, Ontario L8S 4M1, Canada}

\author{Y. Cai}
\affiliation{Department of Physics and Astronomy, McMaster University, Hamilton, Ontario L8S 4M1, Canada}

\author{G.~M.~Luke}
\affiliation{Department of Physics and Astronomy, McMaster University, Hamilton, Ontario L8S 4M1, Canada}
\affiliation{TRIUMF, 4004 Wesbrook Mall, Vancouver, BC V6T 2A3, Canada}

\author{Franziska  Weickert}
\affiliation{National High Magnetic Field Laboratory, Florida State University, Tallahassee, Florida 32310, USA}

\author{Andriy H. Nevidomskyy}
\affiliation{Department of Physics and Astronomy, Rice University, Houston, TX, 77005 USA}

\author{Christos D. Malliakas}
\affiliation{Materials Science Division, Argonne National Laboratory, Lemont, Illinois 60439, USA}

\author{Mercouri G. Kanatzidis}
\affiliation{Materials Science Division, Argonne National Laboratory, Lemont, Illinois 60439, USA}

\author{Shiming Lei}
\affiliation{Department of Physics and Astronomy, Rice University, Houston, TX, 77005 USA}

\author{Kyle Bayliff}
\affiliation{Department of Chemistry, Rice University, Houston, TX, 77005 USA}

\author{E. Morosan}
\affiliation{Department of Physics and Astronomy, Rice University, Houston, TX, 77005 USA}

\date{\today}


\begin{abstract}
New phases of matter emerge at the edge of magnetic instabilities, which can occur in materials with moments that are localized, itinerant or intermediate between these extremes. In local moment systems, such as heavy fermions, the magnetism can be tuned towards a zero-temperature transition at a quantum critical point (QCP) via pressure, chemical doping, and, rarely, magnetic field. By contrast, in itinerant moment systems, QCPs are more rare, and they are induced by pressure or doping, and there are no known examples of field-induced transitions. This means that no universal behaviour has been established across the whole itinerant-to-local moment range, a substantial gap in our knowledge of quantum criticality. Here we report the discovery of a new itinerant antiferromagnet \TiCu{} that can be tuned to a QCP by a small magnetic field. We see signatures of quantum criticality and the associated non-Fermi liquid behaviour in thermodynamic and transport measurements, while band structure calculations point to an orbital-selective, spin density wave ground state, a consequence of the square net structural motif in Ti3Cu4. \TiCu{} thus provides a platform for the comparison and generalisation of quantum critical behaviour across the whole spectrum of magnetism.

\end{abstract}

\maketitle 

\graphicspath{ {Figures/} }

Quantum critical points (QCPs) emerge upon the continuous (second order) suppression of magnetic order to zero temperature via a non-thermal tuning parameters such as doping, pressure or magnetic field. The interplay of magnetism, electron correlations, and quantum critical fluctuations in the vicinity of quantum phase transitions (QPTs) has been linked to novel emergent physics like unconventional superconductivity \cite{johnston2010,keimer2015,dai2012,dai2015antiferromagnetic}, non-Fermi liquid (NFL) behavior \cite{julian1998non,stewart2001non,lohneysen2007fermi,coleman2001fermi}, and heavy fermion behavior \cite{schroder2000onset,custers2003break,gegenwart2008}.

Even though QPTs have been induced by pressure and doping in numerous systems, including local and itinerant magnetic compounds, these tuning parameters present experimental challenges: the former often requires experimentally difficult high pressure values to suppress the transition to $T = 0$, while the latter results in convoluted effects of disorder and quantum criticality, often difficult to resolve separately. Magnetic field appears as an advantageous tuning parameter to study quantum criticality \cite{heuser1998inducement}, although there are much fewer experimental observations of field-induced QCPs. Field induced quantum criticality has been reported in the heavy-fermions YbRh$_2$Si$_2$ \cite{gegenwart2002magnetic,gegenwart2006high,gegenwart2008unconventional}, YbAgGe \cite{tokiwa2013quantum}, CePdAl \cite{zhao2019quantum}, CeCoIn$_5$ \cite{paglione2003field}, CeAuSb$_2$ \cite{balicas2005magnetic}, YbPtIn \cite{morosan2006magnetic}, CePtIn$_4$ \cite{das2019magnetic}, Bose-Einstein condensates (BECs) in quantum magnets \cite{zapf2014bose}, or the metamagnets with either f electrons as in CeRu$_2$Si$_2$ \cite{daou2006continuous} and UCoAl \cite{aoki2011ferromagnetic}, or d electrons in Sr$_3$Ru$_2$O$_7$ \cite{tokiwa2016multiple, grigera2003angular, rost2009entropy,Grigera2001}. No universal behavior can so far be established across the whole itinerant-to-local moment range \cite{hertz1976quantum,millis1993effect,gegenwart2008,millis2002metamagnetic, belitz2017quantum}, in large because of the complexities associated with local moments hybridizing with conduction electrons. It thus seems advantageous to study purely itinerant magnets, \textit{i.e.,} magnetic systems with no partially filled electronic shells. While the only known such itinerant magnets ZrZn$_2$ \cite{matthias1958ferromagnetism,sokolov2006critical}, Sc$_{3.1}$In \cite{matthias1961ferromagnetism, svanidze2015non}, and TiAu \cite{svanidze2015,svanidzeprb2017}, have been tuned to QCPs by doping, the lack of experimental observation of field-induced QCPs in the extreme limit of itinerant moments is likely a reflection of the larger magnetic energy scales associated with d-electron systems compared to their f-electron counterparts. Furthermore, the magnetism in Cr, the prototypical spin-density wave (SDW) system, can be suppressed to a QCP with doping \cite{yeh2002quantum,jaramillo2009breakdown} or pressure \cite{jaramillo2010signatures}, but magnetic field has little or no effect on the ordering temperature \cite{fawcett1988spin}. On the other hand, Sr$_3$Ru$_2$O$_7$ \cite{tokiwa2016multiple, grigera2003angular, rost2009entropy,Grigera2001}, a paramagnet in zero magnetic field, can be tuned to a quantum critical end point (QCEP), where a line of \textit{first-order} itinerant metamagnetic transitions terminates at $T$ = 0, motivating new theories of first-order metamagnetic itinerant quantum criticality \cite{millis2002metamagnetic, belitz2017quantum}. Thus the experimental realization of a field-induced second-order QPT in a purely itinerant magnetic system has until now remained elusive.

Here we report the discovery of the itinerant antiferromagnetic (AFM) metal Ti$_3$Cu$_4$, where Ti and Cu have empty or filled d shells, and therefore neither carry a local moment. The N\'eel temperature $T_N~=$ 11.3 K is continuously suppressed to zero at a magnetic field-induced QCP at a critical field $H_c = 4.87$ T. Concurrently, the magnetic Gr{\"u}neisen ratio $\Gamma_H~=~\frac{1}{T}\frac{\partial {T}}{\partial {H}} \big|_S$ diverges as $H \rightarrow H_c$ and $T\rightarrow0$, with a sign change and divergence in $T$ at $H~=~H_c$, accompanied by a NFL-FL crossover. The continuous suppression of the magnetic order to $T = 0$ by magnetic field, together with the divergence of thermodynamic properties (such as the magnetic Gr{\"u}neisen ratio) are the benchmarks for identifying QPTs. Ti$_3$Cu$_4$ provides a unique platform to study a field-induced QCP at a low field scale for a d-electron itinerant magnet, and without the complexities of the interplay between local and itinerant moments.

\section{Results}

\begin{figure*}[ht]
\includegraphics[scale=0.6]{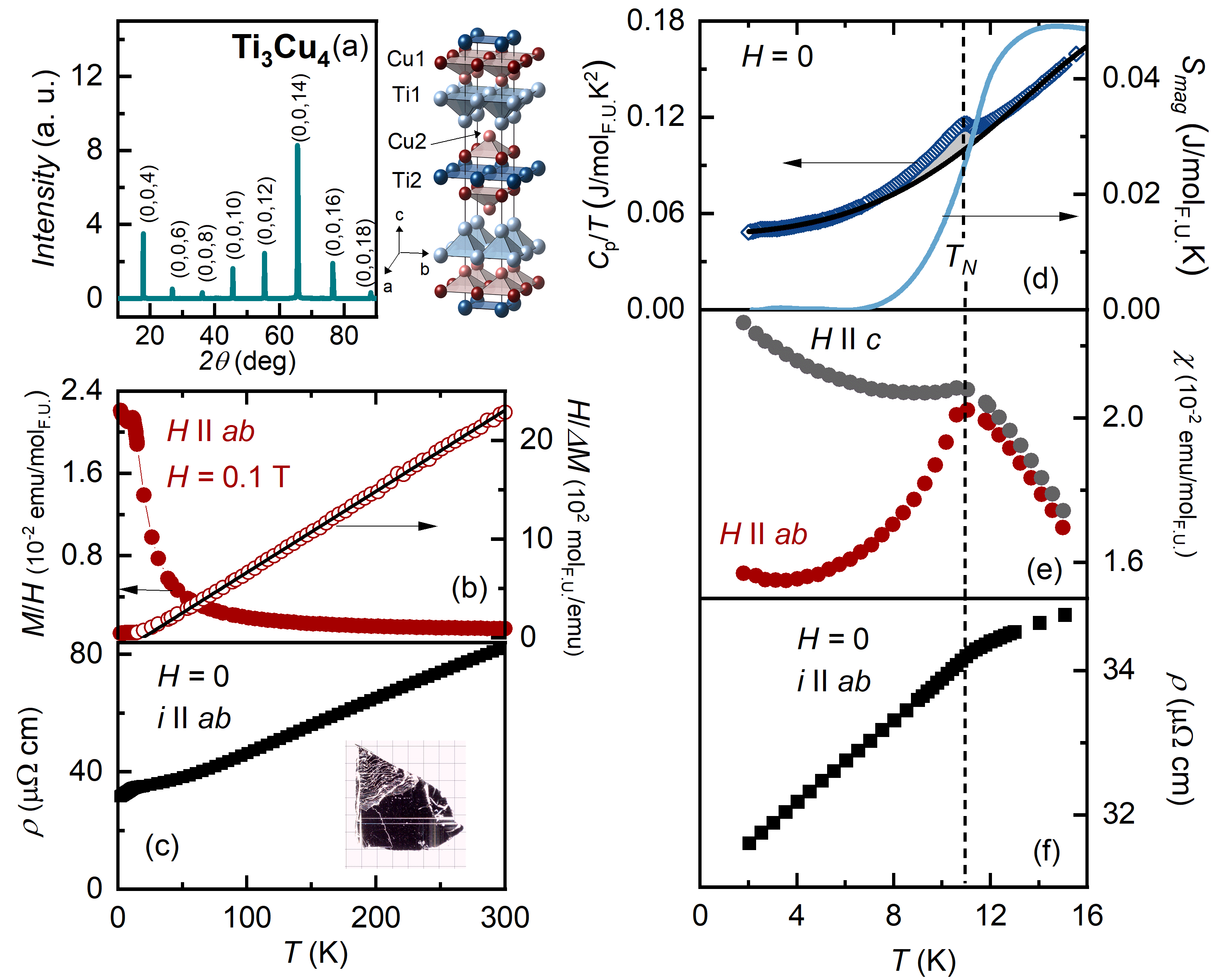}
  \caption{\textbf{Antiferromagnetic order in Ti$_3$Cu$_4$.} (a) X-ray diffraction pattern of single crystalline Ti$_3$Cu$_4$ with the beam incident on the as grown surface showing a series of $(00l)$ Bragg reflections. (right) The tetragonal crystal structure of Ti$_3$Cu$_4$, composed of alternating layers of Ti (blue) and Cu (red), with two unique crystallographic sites for each indicated by the dark and light shading. (b) The magnetic susceptibility (left axis, filled circles) measured in an $H~||~ab = 0.1$~T field, with a cusp at $T_N$ = 11.3 K. The inverse susceptibility (right axis, open circles) is fit with a Curie-Weiss \emph{like} equation (black line) which gives $\mu_{\text{PM}} = 1.0~\mu_B/$F.U. and $T^* = 19.4$~K. (c) Resistivity with current $i~||~ab$ showing a sharp decrease at $T_N$. (inset)  A typical single crystal of Ti$_3$Cu$_4$ with the grid lines spaced 1mm apart.  (d) The heat capacity scaled by temperature (left axis) exhibits a peak at $T_N$. The non-magnetic contribution was fit to a polynomial (black line). The calculated entropy (right axis) saturates at just 0.8\% of $R\ln{2}$. (e) The H$\rightarrow$ 0 magnetic susceptibility  $\chi(T)$ (see text for details), showing an AFM cusp with $H~||~ab$ (red symbols), while $\chi(T)$ plateaus for  $H~||~c$ (grey symbols).  (f) Zoomed in resistivity showing the anomaly at 11.3 K that coincides with the anomalies in susceptibility and heat capacity at $T_N$.}
  \label{fig:F1}
\end{figure*}

Flux-grown single crystals form as flat plates, with typical dimensions of 2~mm $\times$ 2~mm $\times$ 0.5 mm (Fig. \ref{fig:F1}). \TiCu{} crystallizes in the tetragonal $I4/mmm$ space group \cite{schubert1964einige}. The crystal growth and structural characterization details are given in the Methods. X-ray diffraction measurements with the beam incident on the as-grown surface reveal a series of sharp $(00l)$ Bragg reflections shown in Fig. \ref{fig:F1}a, consistent with the $I4/mmm$ symmetry. This layered structure shown in Fig. \ref{fig:F1}a, right, contains two different crystallographic sites for both Cu (light and dark red) and Ti (light and dark blue). Alternating layers of Ti are arranged in buckled (Ti1) and square (Ti2) nets, separated by staggered buckled nets of Cu. The connectivity of the Ti2 atoms in \TiCu{} is likely responsible for its remarkable electronic and magnetic properties, as discussed below. 

The DC magnetic susceptibility $M(T)/H$ for $H = 0.1$~T (Fig.~\ref{fig:F1}b, full symbols, left axis), shows Curie-Weiss-\emph{like} temperature dependence, with no irreversibility in the field-cooled (FC) and zero-field-cooled (ZFC) data. Throughout the paper, only ZFC data is shown for clarity.  Indeed, the inverse susceptibility $H/\Delta M$ is linear in $T$ down to $\sim$ 20 K (Fig.~\ref{fig:F1}b, open symbols, right axis), where $\Delta M= M-M_0$ corresponds to the magnetization after a small temperature-independent Pauli term, $M_0$ = 4.5 $\times$ 10$^{-4}$ emu/mol$_\mathrm{F.U.}$, has been subtracted. In the same temperature range, $H$ = 0 resistivity $\rho(T)$ measurements (Fig. \ref{fig:F1}c) reveal the metallic character of \TiCu{}, as  $\rho(T)$ decreases monotonically with decreasing \textit{T}, before a drop at the lowest temperatures. Together, these two measurements provide preliminary indication of itinerant moment magnetism in \TiCu{}, which will be more convincingly demonstrated once the nature of the low temperature phase transition is established.

The low temperature thermodynamic and transport data show that a phase transition in Ti$_3$Cu$_4$ occurs around 11 K (Fig. \ref{fig:F1}d-f), first revealed by the small peak in the specific heat scaled by temperature $C_p/T$ (symbols, Fig. \ref{fig:F1}d, left axis). While such a transition could have a structural component, this is ruled out by single crystal neutron diffraction experiments (discussed below) that show no detectable change to the crystal structure down to 5 K. The antiferromagnetic order at $T_N$ = 11.3 K is confirmed by the $H$ = 0 susceptibility $\chi(T)$ and electrical resistivity $\rho(T)$ (Fig.~\ref{fig:F1}e-f). Anisotropic $\chi(T)$ data (determined from low $H$ magnetization isotherms $M(H)$, as described in the Supplementary Materials (Fig. \ref{fig:SF1}) reveal a peak at $T_N$ for $H\parallel ab$ (red symbols, Fig.~\ref{fig:F1}e), and a nearly temperature-independent plateau below $T_N$ for $H\parallel c$ (grey symbols, Fig.~\ref{fig:F1}e). In a local moment picture, such magnetic anisotropy would be consistent with an AFM ordered state; the susceptibility peaks at $T_N$ when the field is parallel to the direction of the ordered moments. The implication for the itinerant AFM order in \TiCu{} is that the moments are likely oriented within the $ab$ plane, consistent with the single crystal elastic neutron scattering experiments discussed later. Upon cooling through $T_N$, a drop in resistivity signals loss of spin-disorder scattering (full symbols, Fig. \ref{fig:F1}f), with a peak in the resistivity derivative $d\rho/dT$, coincident with the peak in C$_p/T$ and susceptibility derivative $d(\chi T)/dT$ (Fig. \ref{fig:SF2}) \cite{fisher1962relation, fisher1968resistive}. 

\begin{figure}
 \includegraphics[width=\columnwidth]{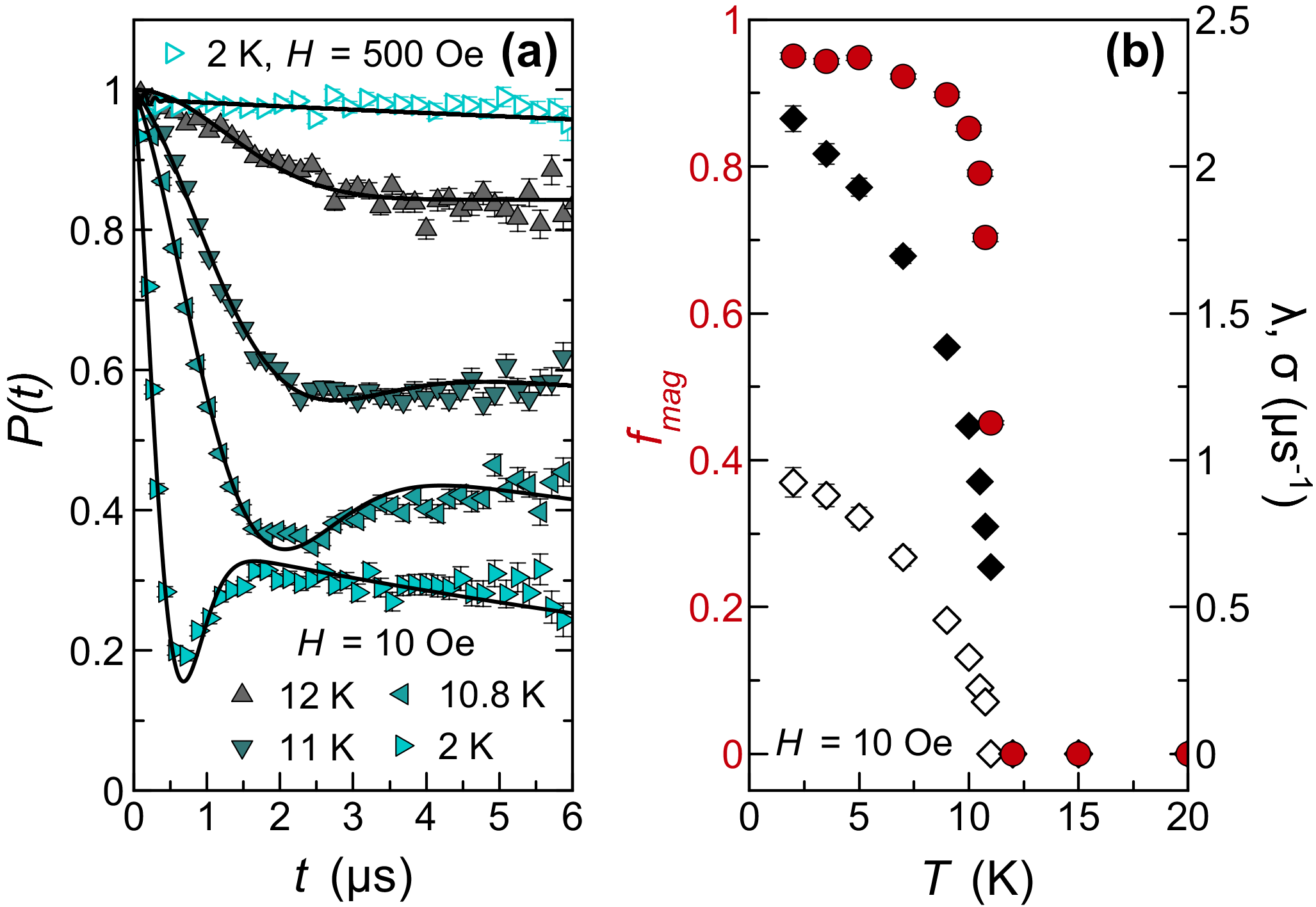}
  \caption{\textbf{Bulk magnetism in Ti$_3$Cu$_4$ from muon spin relaxation.} (a) Representative muon decay asymmetry for Ti$_3$Cu$_4$ at $H=10$ Oe for various temperatures (filled triangles) as well as at $T=2$~K and $H = 500$ Oe (open triangles) with fits to Eqn.~1 (solid lines), showing the onset of static magnetic order. (b) The temperature dependence of the fit parameters: the magnetic volume fraction, $f_{mag}$ (red circles, left hand axis) and the Gaussian, $\sigma$ (filled diamonds) and Lorentzian, $\lambda$ (open diamonds) relaxation rates.} 
  \label{fig:F3}
\end{figure}

Muon spin relaxation ($\mu$SR) measurements were performed, in order to confirm that the magnetic order at $T_N = 11.3$ K in Ti$_3$Cu$_4$ is intrinsic, and not arising from a small impurity phase. Several representative muon decay asymmetry spectra $P(t)$ are plotted in Fig.~\ref{fig:F3}a. A small $H = 10$ Oe field was applied to decouple any relaxation due to nuclear dipoles. From 12 to 20 K, $P(t)$ is temperature independent and exhibits slow relaxation, consistent with a paramagnetic state. Upon cooling through $T_N = 11.3$ K, there is a sharp increase in the relaxation at early times. Within the magnetically ordered state, $P(t)$ takes a characteristic Kubo-Toyabe form \cite{hayano1979zero} with a minimum at early times followed by a recovery to 1/3 of the initial asymmetry. The solid lines in Fig.~\ref{fig:F3}a are fits to $P(t)$ of the following form:
\begin{align}
P(t)  = & (1-f_{mag}) \cdot e^{-a t} \\ 
+ & f_{mag}\left(\frac{1}{3}+\frac{2}{3}\cdot(1-\sigma^2t^2-\lambda t) \cdot e^{-\frac{\sigma^2t^2}{2}-\lambda t}\right)\cdot e^{-b t}. \nonumber
\label{muSRfit}
\end{align}
Muons that land in the non-magnetic fraction of the sample, $1-f_{mag}$, experience a weak temperature-independent exponential relaxation. The magnetic fraction of the sample, $f_{mag}$, is well-described by a combined Kubo-Toyabe function, where the Gaussian relaxation is given by $\sigma$ and the Lorentzian relaxation by $\lambda$. The dynamics in the 1/3 tail are phenomenologically captured by the inclusion of an exponential relaxation. The temperature dependence of the fitted parameters, $f_{mag}$, $\sigma$, and $\lambda$, is presented in Fig.~\ref{fig:F3}b, where each is observed to sharply increase below $T_N = 11.3$ K. At the lowest temperatures, $f_{mag}$ (full circles, left axis) is close to 100\%, confirming that the magnetsim in Ti$_3$Cu$_4$ is an intrinsic bulk property. The static nature of the magnetic order is confirmed through longitudinal field $\mu$SR measurements, where the relaxation is significantly decoupled by fields as small as $H=50$ Oe and fully decoupled by a field of $H = 500$ Oe (open triangles, Fig.~\ref{fig:F3}a).

With $\mu$SR measurements confirming the intrinsic magnetism, we performed single crystal elastic neutron scattering measurements to investigate the nature of the magnetically ordered state in Ti$_3$Cu$_4$. Measurements above ($T = 20$~K) and below ($T=5$~K) $T_N$ reveal the formation of magnetic Bragg peaks on several high symmetry positions, including (100) and (001), as shown in the rocking curve scans in Fig.~\ref{OP}a,b (for all measured reflections see Supplemental Material Fig. \ref{BraggPeaks}). The double peak that appears for (001) and the other reflections with non-zero $l$ component are not intrinsic, but rather the result of two closely-aligned grains. The intensity of the (100) and (001) Bragg peaks (measured both on warming and cooling) as a function of temperature is presented in Fig.~\ref{OP}c, confirming that the onset of magnetic order occurs at $T_N = 11.3$~K without measurable hysteresis. While the (001) and (100) Bragg peaks were measured in different sample geometries, and therefore their intensities cannot be directly compared, it is nonetheless evident that (001) is significantly more intense than (100), indicative of ordered moments that lie in the $ab$-plane, consistent with the low field susceptibility.

The commensurate positions where magnetic Bragg peaks form in Ti$_3$Cu$_4$ are not allowed by the body-centered selection rules ($h+k+l = 2n$) for the $I4/mmm$ structure and therefore no nuclear Bragg peaks are observed on these positions. We can index these magnetic Bragg reflections with a propagation vector of $\textbf{k} = (0~0~1)$. We proceed by assuming that, as indicated by the DFT calculations discussed later, the magnetism in Ti$_3$Cu$_4$ originates from the conduction bands of the Ti atoms which occupy the $2b$ Wyckoff site (Ti2). It should be emphasized that the neutron data cannot independently distinguish which of the atomic sites in Ti$_3$Cu$_4$ is responsible for the magnetism. There are two symmetry-allowed irreducible representations for the $2b$ Wyckoff site with a $\textbf{k} = (0~0~1)$ propagation vector within the $I4/mmm$ space group: $\Gamma_3$ ($c$-axis antiferromagnet) and $\Gamma_9$ ($ab$-plane antiferromagnet). While both of these magnetic structures produce Bragg peaks at (100), only $\Gamma_9$ yields intense reflections at (001) and (003), consistent with our experiment. The periodicity of this structure is shown in Fig.~\ref{OP}d, consistent with a transverse commensurate spin density wave order. Linear combinations of the two basis vectors that make up $\Gamma_9$ allow a continuous rotation within the $ab$-plane and we cannot determine the exact moment orientation in an unpolarized neutron experiment. The magnitude of the ordered moment, which was estimated by comparing the intensity of the nuclear and magnetic reflections in the ($h0l$) plane measurements and assuming a Ti$^{3+}$ magnetic form factor, is 0.17(5)~$\mu_B$, in good agreement with the high field magnetization data discussed next.

\begin{figure}[tbp]
\linespread{1}
\par
\includegraphics[width=\columnwidth]{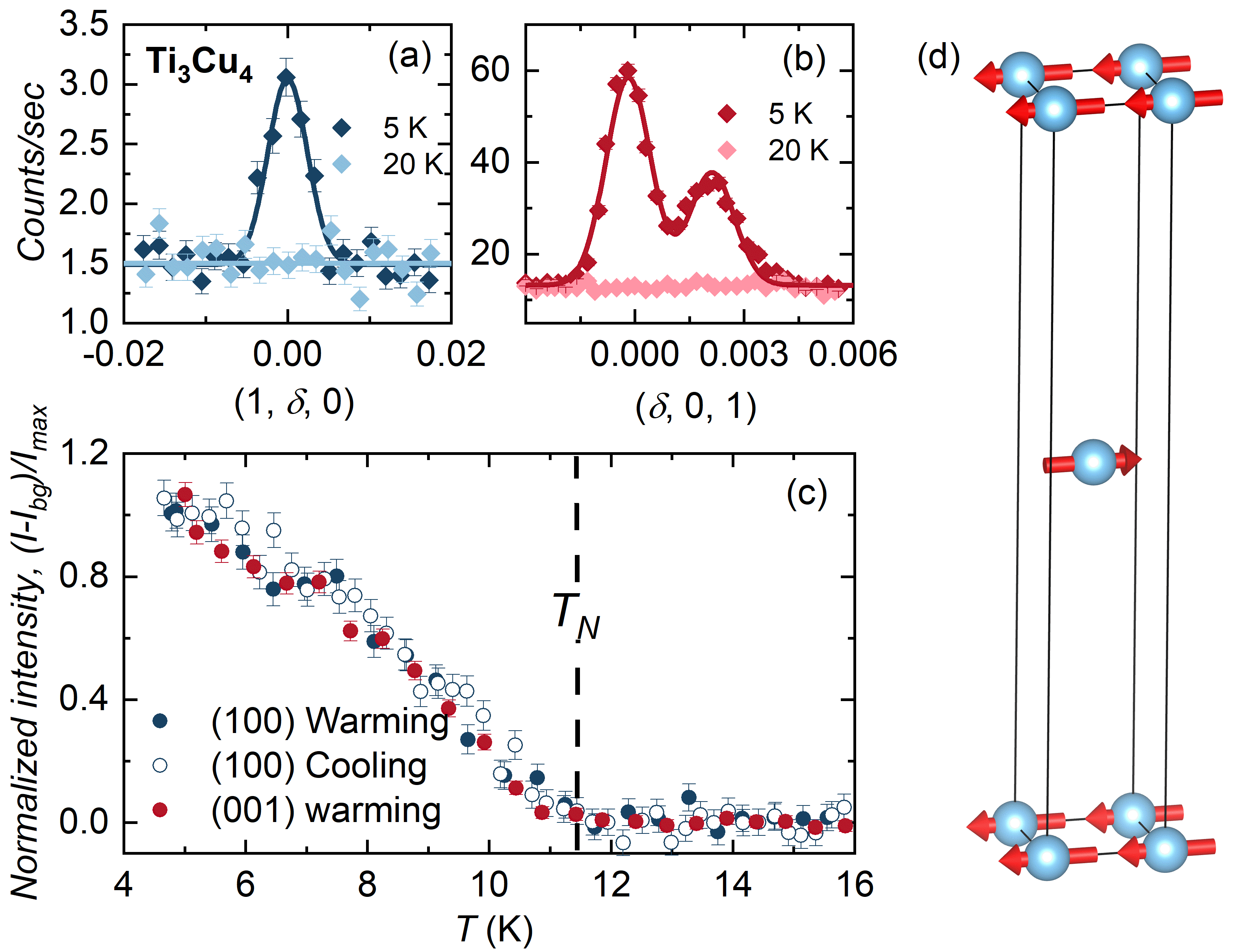}
\par
\caption{\textbf{Elastic neutron scattering.} Rocking curve measurements on the (a) (100) and (b) (001) positions at $T = 20$ and 5~K reveal the formation of magnetic Bragg peaks. Solid lines are fits to a Gaussian lineshape. Note that the maximum divergence in the orthogonal direction for the rocking curves was of order 0.02\%. (c) An order parameter, constructed by measuring the intensity of the (001) and (100) Bragg peaks as a function of temperature, confirms that the magnetic order onsets at $T_N=11.3$~K. (d) The periodicity of the magnetic order as determined by symmetry analysis for Ti2 at the $2b$ Wyckoff site in the $I4/mmm$ space group.}
\label{OP}
\end{figure}

With the bulk antiferromagnetic magnetic order below $T_N$ firmly established, we turn to further evidence of itinerant moment magnetism in \TiCu{}. Recalling the linear inverse susceptibility of \TiCu{} (Fig. \ref{fig:F1}b), we recognize it as signature of either local $\emph{or}$ itinerant moment magnetism, albeit with very different origins. For the former case, mean field theory predicts $\chi(T) \sim \frac{\mu^2_{PM}}{(T - \theta)}$, where $\mu_{PM}$  is the paramagnetic moment, and $\theta$ is a measure of the \emph{inter}-atomic moment coupling. For the latter case, Moriya's theory of spin fluctuations \cite{moriya1973,hasegawa1974,takahashi1985,nakayama1987,konno1987,moriya1981} predicts $\chi(T) \sim \frac{I}{(T - T^*)}$, where $I$ is a measure of the \emph{intra}-atomic coupling. In Ti$_3$Cu$_4$, the slope and intercept of the linear fit to $H/\Delta M$ between 50 and 300 K (solid line, Fig. ~\ref{fig:F1}b) yield a paramagnetic moment $\mu_{\text{PM}}$ = 1.0~$\mu_{B}/$F.U. and $T^*~=~19.4$ K, respectively, where $T^*$ is analogous to the Weiss temperature in local moment systems. The positive $T^*$ is consistent with ferromagnetic in-plane interactions characteristic of the $\Gamma_9$ magnetic structure, where the $c$ direction coupling is AFM. 

While the linear inverse susceptibility alone is not enough to indicate itinerant moments in \TiCu{}, the paramagnetic moment $\mu_{PM}$ is too small to be explained by a local moment scenario, in which the smallest possible unscreened moment would be 1.73 $\mu_{B}/$F.U. corresponding to  $S=\sfrac{1}{2}$ at the Ti2 site (all other sites in this structure have higher multiplicities and would therefore produce even larger magnetic moments per F.U.).

The magnetic entropy S$_{mag}$ (estimated from the grey area under the $C_p/T$ peak in Fig. \ref{fig:F1}d) falls in line with the same conclusion: S$_{mag}$ (thin line, right axis in Fig. \ref{fig:F1}d) reaches only $\sim$ 1\% Rln2 up to 16 K (above $T_N$). Such small entropy release is consistent with small moment ordering, likely smaller even than that in the itinerant antiferromagnet TiAu \cite{svanidze2015} where $S_{mag}$ was close to 3\% $R\ln{2}$. This indicates that the paramagnetic moment in Ti$_3$Cu$_4$ is best explained as originating from itinerant spin fluctuations, a scenario corroborated below by our \textit{ab initio} calculations.

\begin{figure}
 \includegraphics[width=\columnwidth]{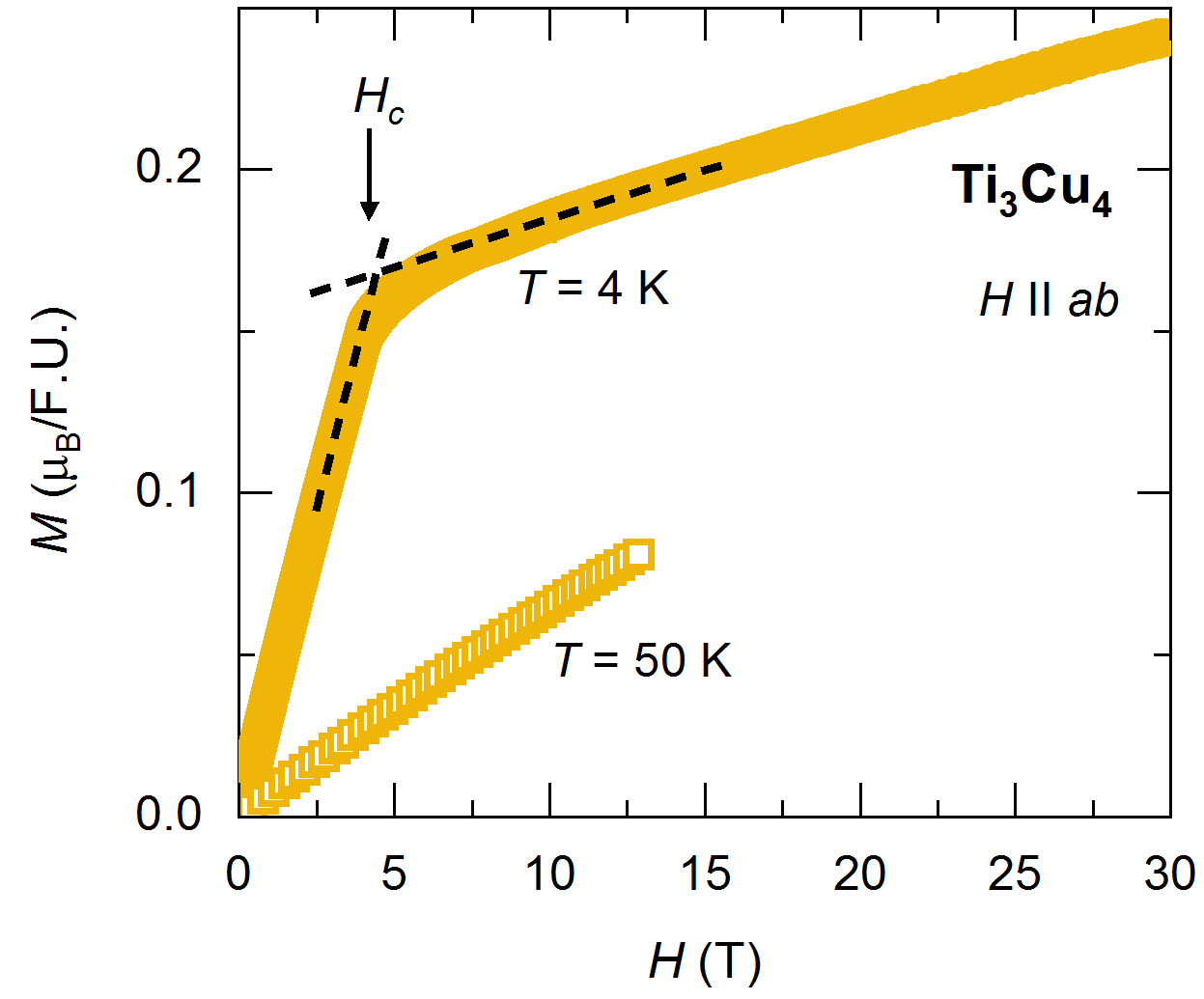}
  \caption{\textbf{Field dependence of the magnetization in Ti$_3$Cu$_4$.} M(H) isotherms measured at $T = 50$~K (yellow, open squares), and  $T = 4$~K (yellow, solid squares) . Lines were fit above and below the metamagnetic transition near 4~T (black, dashed lines). The intersection gives the critical field $H_C$ = 4.3 T at 4K.} 
  \label{fig:F2}
\end{figure}

Another empirical signature of itinerant moment magnetism is a divergent Rhodes-Wohlfarth ratio $q_c/q_s~>>$ 1, where $q_c$ and $q_s$ correspond to the number of magnetic carriers above and below the ordering temperature \cite{rhodes1963effective}. Experimentally, $q_c$ is extracted from the paramagnetic moment $\mu_{PM}$ determined from high temperature fits of the inverse magnetic susceptibility:
\begin{equation}
    \mu_{PM}^2 = q_c(q_c+2)\mu_B^2, \mbox{~or~} q_c \sim 0.4,
\end{equation}

and $q_s$ is determined from the low temperature (ordered) moment $\mu_{ord}$ 
\begin{equation}
    \mu_{ord}  = q_s\mu_B,
\end{equation}
The Rhodes-Wohlfarth ratio $q_c/q_s$ close to unity corresponds to the local moment scenario, while an increase in $q_c/q_s$ with lower ordering temperature indicates an increased degree of itinerancy \cite{rhodes1963effective}. Magnetization measurements $M$($H$) for Ti$_3$Cu$_4$ (Fig.~\ref{fig:F2}) point to a small $\mu_{7T} \sim 0.25~ \mu_B$, while single crystal neutron measurements indicate that the ordered moment $\mu_{ord}$ is even smaller, $0.17(5)~\mu_B$. These values result in a large $q_c/q_s$ $\approx$ 2.4, reinforcing the itinerant magnetism picture in \TiCu{}.

\begin{figure}[t!]

 \includegraphics[width=\columnwidth]{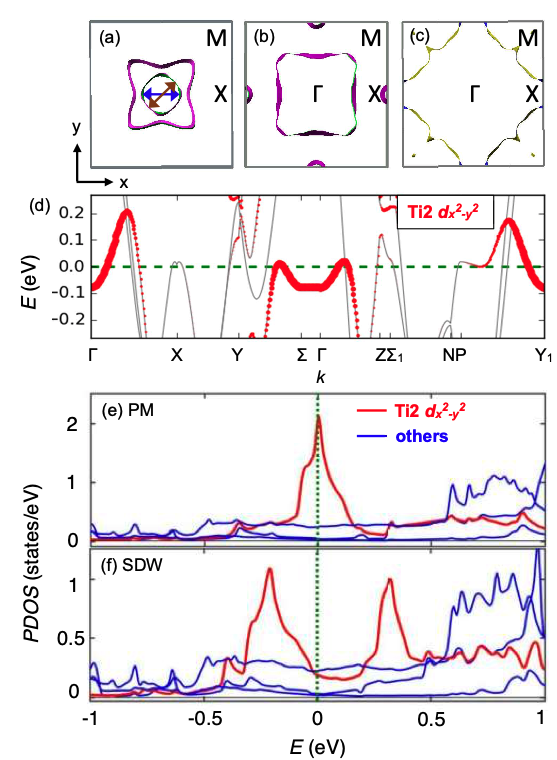}
 \caption{\textbf{Electronic structure from DFT calculation.} (a), (b) and (c) Fermi sheets constituting the Fermi surface (FS) of  \TiCu{}. Panel (a) displays the FS originating from the $d_{x^2-y^2}$ orbital of Ti2, which contributes most to the DOS at the Fermi level, see panel (e). The arrows indicate possible nesting wave-vectors of the Fermi surface. (d) Band structure near the Fermi level of  \TiCu{}. The width of the red line is proportional to the projection onto the $d_{x^2-y^2}$ orbital of Ti2. (e) and (f) Projected density of states in the paramagnetic (PM) and SDW phase, respectively, with the red line indicating the contribution of Ti2 $d_{x^2-y^2}$ orbital. The blue lines indicate the partial density of states of the other Ti2 $d$-orbitals, which are comparatively negligible at the Fermi level.}
 \label{fig:theo_main_fig}
  \end{figure}

In order to glean insight into the nature of the magnetic order, and in particular the small value of the ordered moment, we performed first principles calculations based on density functional theory (DFT), with the methodology detailed in Supplementary Materials. The calculations reveal a Fermi surface consisting of four sheets centered around the $\Gamma$ point, and a small pocket around the $X$ point (Fig.~\ref{fig:theo_main_fig}a-c). 
The analysis of the orbital-projected band structure (so-called ``fat bands") in Fig.~\ref{fig:theo_main_fig}d shows that the main contribution to the nested Fermi surface sheet in Fig.~\ref{fig:theo_main_fig}a comes from the $d_{x^2-y^2}$ orbital on the Ti2 atom, whereas the partial contributions from the other orbitals and from Ti1 atoms are much smaller, as demonstrated by the projected density of states (DOS) in Fig.~\ref{fig:theo_main_fig}e. The reason for this orbital selectivity appears to be connected to the square net geometry of the Ti2 layer, where the $d_{x^2-y^2}$ orbital lies along the Ti2-Ti2 bonds, reminiscent of the cuprates \cite{comin2016resonant}.

 \begin{figure}
 \includegraphics[width=\columnwidth]{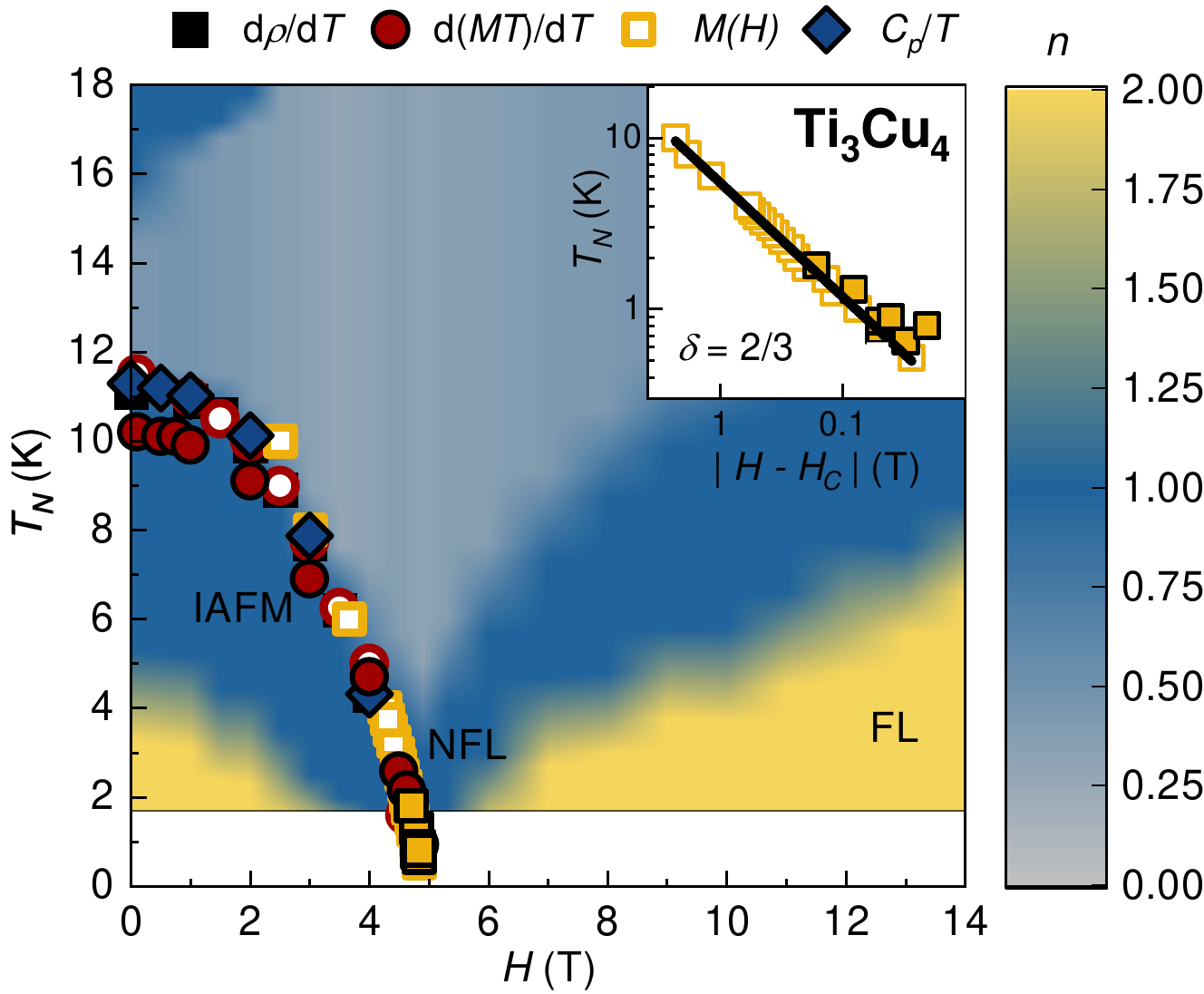}
  \caption{\textbf{H - T phase diagram for Ti$_3$Cu$_4$.} The AFM ordering temperature and field were determined from $d(MT)$/d$T$ (red circles), $M(H)$ (yellow squares), $C_p/T$ (blue diamonds), and $\text{d}\rho/\text{d}T$ (black squares) Closed and open symbols denote measurements with $H \parallel c$ and $H \parallel ab$, respectively. The itinerant AFM order is fully suppressed for fields above $H_C=4.865$~T, and the contour plot maps the resistivity exponent $n$ from fits of $\rho~=~\rho_0~+~A_nT^{n}$, exhibiting a crossover from non-Fermi liquid behavior ($n<2$) to a Fermi liquid region $n = 2$ as the QCP is crossed in the field direction at the lowest measured temperatures. Below 1.8~K, the white region corresponds to temperatures not accessed by our $\rho(T)$ experiments. (inset) A log-log plot of $T_N$ vs. $|H-H_C|$ (yellow squares), with the black line corresponding to the fit of $T_N \propto |H-H_C|^\delta$, yielding $H_C = 4.87$ $\pm 0.005$ T and $\delta$ ~=~2/3.}    
 
  \label{fig:F4}
\end{figure}

The calculations performed in the magnetically ordered phase, with the experimentally determined wavevector $\mathbf{k}=(0~0~1)$  show that the DOS gets depleted around the chemical potential (Fig.~\ref{fig:theo_main_fig}f), and that the sharp DOS peak present in the PM phase (Fig.~\ref{fig:theo_main_fig}e) is split into two peaks separated by about 0.5~eV, with a pseudogap in between. This peak separation, due to the internal staggered magnetic field, is quantitatively consistent with the DFT-predicted ordered moment of 0.25~$\mu_B$ per Ti2 ion. Interestingly, the calculations show zero ordered moment on the buckled Ti1 layer. The reason is that the center of mass of the Ti1 $d_{x^2-y^2}$ band lies far below the Fermi level (close to $-1$~eV) due to greater hybridization with the $d_{xz}$ and $d_{yz}$ orbitals within the buckled layer, thus unable to participate in the formation of the magnetic order on the Ti1 sites.

The above analysis, combined with the smallness of the magnetic moment on the Ti2 ion, clearly indicates the itinerant nature of the magnetism in  \TiCu{}. Of note, the Fermi sheets in Fig.~\ref{fig:theo_main_fig}a-c appear to be nested, suggesting that a spin-density wave (SDW) order is likely to be realized with wavevectors along either $\mathbf{k}_1 \parallel (1~0~0)$ (brown arrow in panel a) or $\mathbf{k}_2 \parallel (~1~1~0)$ (blue arrow). However, the neutron diffraction instead shows an out-of-plane wave-vector $\mathbf{k}=$ $(0~0~1)$. In order to elucidate this puzzling behaviour, we performed a series of \textit{ab initio} calculations with various ordering wavevectors, as described in the Supplementary Materials. Figure ~\ref{fig:Supp-sdw} shows that the candidate SDW states with various commensurate wave-vectors along $(1~0~0)$ and $(1~1~0)$ are all higher in energy than that of the experimentally observed $\mathbf{k}=(~0~0~1)$ state, with one notable exception: the noncollinear $(\frac{1}{8}~0~0)$ state is predicted to lie slightly (about 1 meV/f.u.) lower in energy. This energy difference is however within the error bars of the DFT calculation and is therefore not significant. We conclude that the nested nature of the Fermi surfaces allows for several candidate SDW states very close in energy. We therefore rely on the neutron diffraction study to deduce the ordered state with the wavevector $\mathbf{k}=(0~0~1)$.

We return to the field-induced transition in \TiCu{}. Increasing magnetic field continuously suppresses $T_N$ as seen in $M(H)$, $M(T)/H$, $C_p$, and $\rho(T)$ measurements for both $H \parallel ab$ and $H \parallel c$ (Supplementary Materials Fig.~\ref{fig:SF3} and \ref{fig:Res}). A field $H_c~\sim$ 4.87 T suppresses the magnetic order all the way to zero temperature, as shown in the $T-H$  phase diagram in Fig.~\ref{fig:F4} and \ref{fig:Res}, raising the possibility of a field-induced QCP in \TiCu{}. Down to 0.5 K, the transition is continuous, with no apparent hysteresis. The log-log $T-H$ plot around $H_C$ is linear (inset, Fig. \ref{fig:F4}),  such that the phase boundary in the vicinity of $H_C$ can be described by an exponential behavior \textit{T} $\propto~|H~-~H_C|^{\delta}$, with $H_C$ = 4.87 $\pm 0.005~$ and $\delta$ = $2/3$. This corresponds to the expected Hertz-Millis exponent for a 3D AFM \cite{hertz1976quantum,millis1993effect, stewart2001non} or Bose-Einstein condensation of magnons \cite{giamarchi1999coupled,nikuni2000BEC, nohadani2004universal}. 

Thermodynamic measurements provide convincing evidence of the field-induced quantum criticality \cite{rost2009entropy,gen2019magnetocaloric,tokiwa2009divergence}, so in Ti$_3$Cu$_4$, we turn to the magnetic Gr{\"u}neisen ratio $\Gamma_H$ defined as \cite{zhu2003universally}

\begin{equation}
    \Gamma_H = -\frac{(\partial {M}/\partial {T})_H}{C_H} =-\frac{(\partial {S}/\partial {H})_T}{T(\partial {S}/\partial {T})_H} = \frac{1}{T}\frac{\partial {T}}{\partial{H}}\bigg|_{S},
    \label{MCEeq}
\end{equation}

\noindent which measures the slope of the isentropes of the magnetic phase boundary in the $H~-~T$ plane \cite{garst2005sign}. Across a classical phase transition, $\Gamma_H$ is expected to be finite and temperature-independent \cite{zhu2003universally}. Near a QCP, an entropy ridge is expected to form where the system is maximally undecided between the ordered state and the disordered state (for d$H~>~0$, ($\partial {S}/\partial {H})_T~>~0$ when $H~<~H_c$ and ($\partial {S}/\partial {H})_T~<~0$ when $H~>~H_c$), which is reflected by a sign change of $\Gamma_H$ at the QCP \cite{garst2005sign}.  Furthermore, in the low-temperature limit, the singularities in $S$ and $T$ cancel out in Eq.~\ref{MCEeq}, leaving only singularities associated with $H$ \cite{zhu2003universally}. Zhu \textit{et al.} showed that $\Gamma_H$ scales as $\frac{1}{H-H_c}$ as $T\rightarrow 0$ and $\Gamma_H\sim \frac{1}{T^{1/\nu z}}$ at $H~=~ H_c$ \cite{zhu2003universally}. Here $\nu$ is the exponent of the correlation length and $z$ is the dynamical critical exponent. Together, the sign change of $\Gamma_H$ at $H_c$ and the scaling relations are definitive proof of a field induced QCP \cite{garst2005sign,zhu2003universally,gegenwart2016gruneisen}.

 \begin{figure}[t]
 \includegraphics[width=\columnwidth]{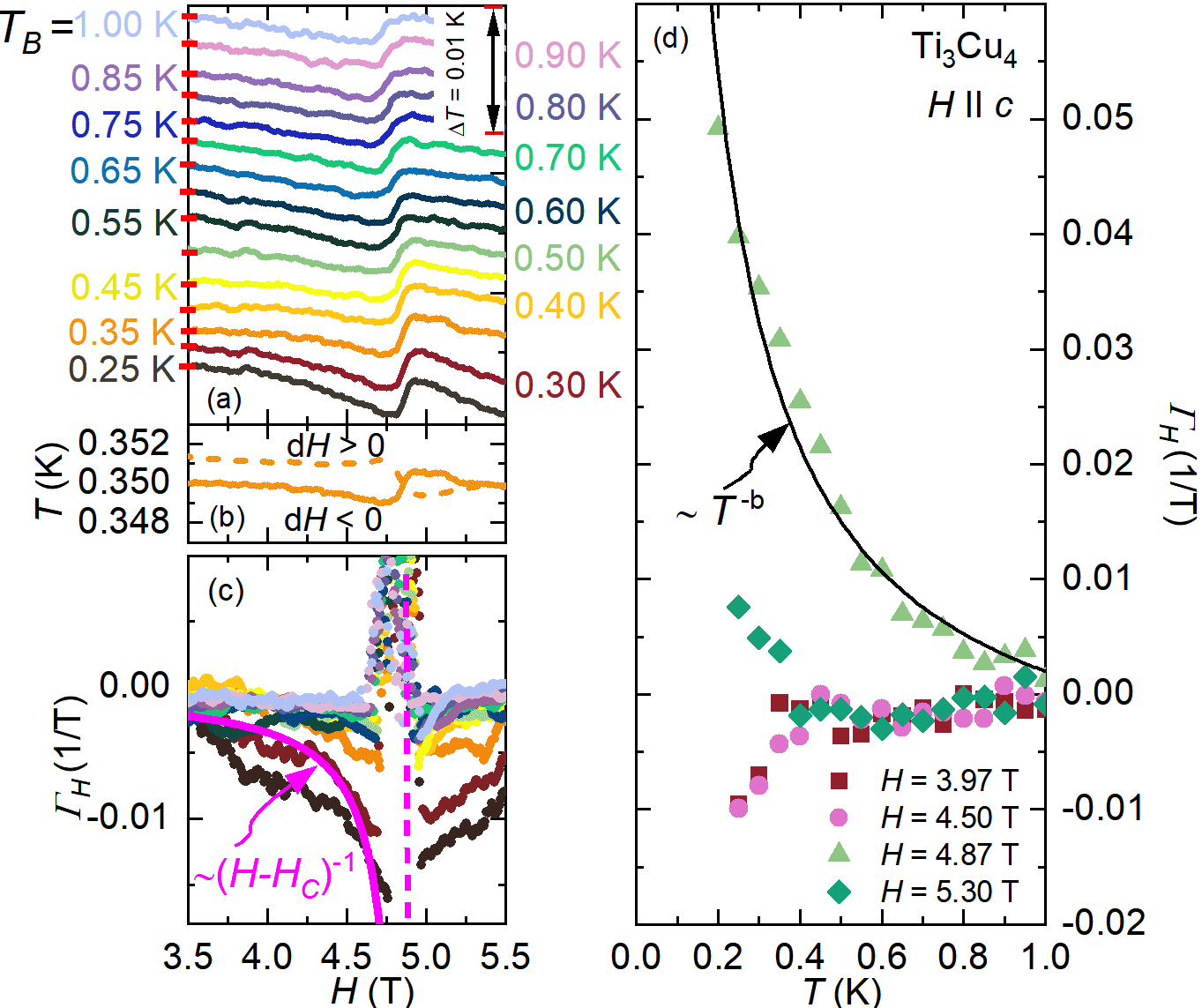}
  \caption{\textbf{Divergence of MCE at the critical field}. (a) MCE measurements for $ H \parallel c$ and d$H > 0$ measured at various bath temperatures $T_B$. Data are offset by arbitrary values for clarity, where the red lines indicated a zero change in temperature $\Delta T$, and the scale bar on the right gives the absolute change in temperature. (b) A zoom in of the $T_B = 0.35~$K MCE data measured with d$H > 0$ (solid line) and d$H < 0$ (dashed line). (c) The field-dependent magnetic Gr{\"u}neisen ratio (see text for details) divergences as $H$ approaches $H_C$. The pink solid dashed line is a guide to the eye and is proportional to $\frac{1}{H-H_c}$, while the vertical pink dashed  line denotes $H_C$. (d) The temperature dependence of the magnetic Gr{\"u}neisen ratio at selected fields below $H_C$ (red squares and pink circles ), at $H_C$ (light green triangles), and above $H_C$ (dark green diamonds). At $H = H_C$, the magnetic Gr{\"u}neisen ratio diverges as $T^{-b}$ (solid black line). } 
 
  \label{fig:MCE}
\end{figure}

We obtain $\Gamma_H$ by measuring the magnetocaloric effect (MCE) ($\partial {T}/\partial{H}$) under quasi-adiabatic conditions ($S\sim$ constant for a duration smaller than it takes for the thermometer to relax). In Fig.~\ref{fig:MCE}a, we plot MCE, \textit{i.e.} the temperature change driven by ramping the magnetic field $H$ across $H_C$ from 3.5~T to 5.5~T at various bath temperatures $0.25~K<T_B<1.00~K$ for $H \parallel c$. Upon increasing $H$ from below H$_c$, $T$ decreases, such that $\partial T/\partial H~<~0$. Since $C_H$ is a positive quantity, the sign of $\partial T/\partial H$ is always opposite to the sign of $\partial S/\partial H$. Consequently, the decrease in $T$ during the field upsweep indicates an increase in magnetic entropy ($\partial S/\partial H~>0$). Near $H_C$, there is a sudden increase in $T$, indicating a sudden reduction of the magnetic entropy. Subsequently, $T$ decreases again due to the measurement apparatus relaxing back to $T_B$. To confirm that the decrease in $T$ above $H_C$ is indeed related to the measurement apparatus relaxation and not intrinsic to the sample, we measured the MCE sweeping $H$ down from 5.5~T to 3.5~T (Fig.~\ref{fig:MCE}b, dashed line). Upon decreasing H above $H_C$, $T$ decreases indicating an increase in magnetic entropy as the QCP is approached. The sudden increase in $T$ reflects a decrease in magnetic entropy as $H$ crosses $H_c$. Upon further decreasing $H$, the temperature again relaxes towards $T_B$ before increasing due to a reduction of magnetic entropy as the distance from $H_c$ is increased.

Figure~\ref{fig:MCE}c shows $\Gamma_H(H)$ at selected $T_B$, approximated as $\Gamma_H~\approx~ \frac{1}{T_B}\frac{\Delta T}{\Delta H}$. Though we cannot reliably extract the exponents due to the quasi-adiabatic nature of our experiments, as $T~\rightarrow~0$ it is apparent that $\Gamma_H$ diverges as $(H-H_c)^{-1}$, as illustrated by the pink solid line. Furthermore, $\Gamma_H(T)$ (Fig. \ref{fig:MCE}d) switches signs across $H_C$ and diverges as $T^{-b}$ (black solid line) for $H~=~H_C$. While the quasi-adiabiatic conditions render the exponents' determination uncertain, the MCE power law divergence is unambiguous: assuming constant heat loss, the exponent may vary, but such a scenario cannot cause a divergence. For a classical phase transition, the Gr{\"u}neisen ratio is a constant, and therefore the divergence must come from the QCP.  
Together, the sign change of $\Gamma_H$ across $H_C$ and the divergences at $\Gamma_H(H,T\rightarrow 0)$ and $\Gamma_H(H=H_C,T)$ provide ample evidence for a field induced QCP at $H_C~=~4.87~T$ \cite{zhu2003universally,garst2005sign}.

\begin{figure*}[ht]
\includegraphics[scale=0.5]{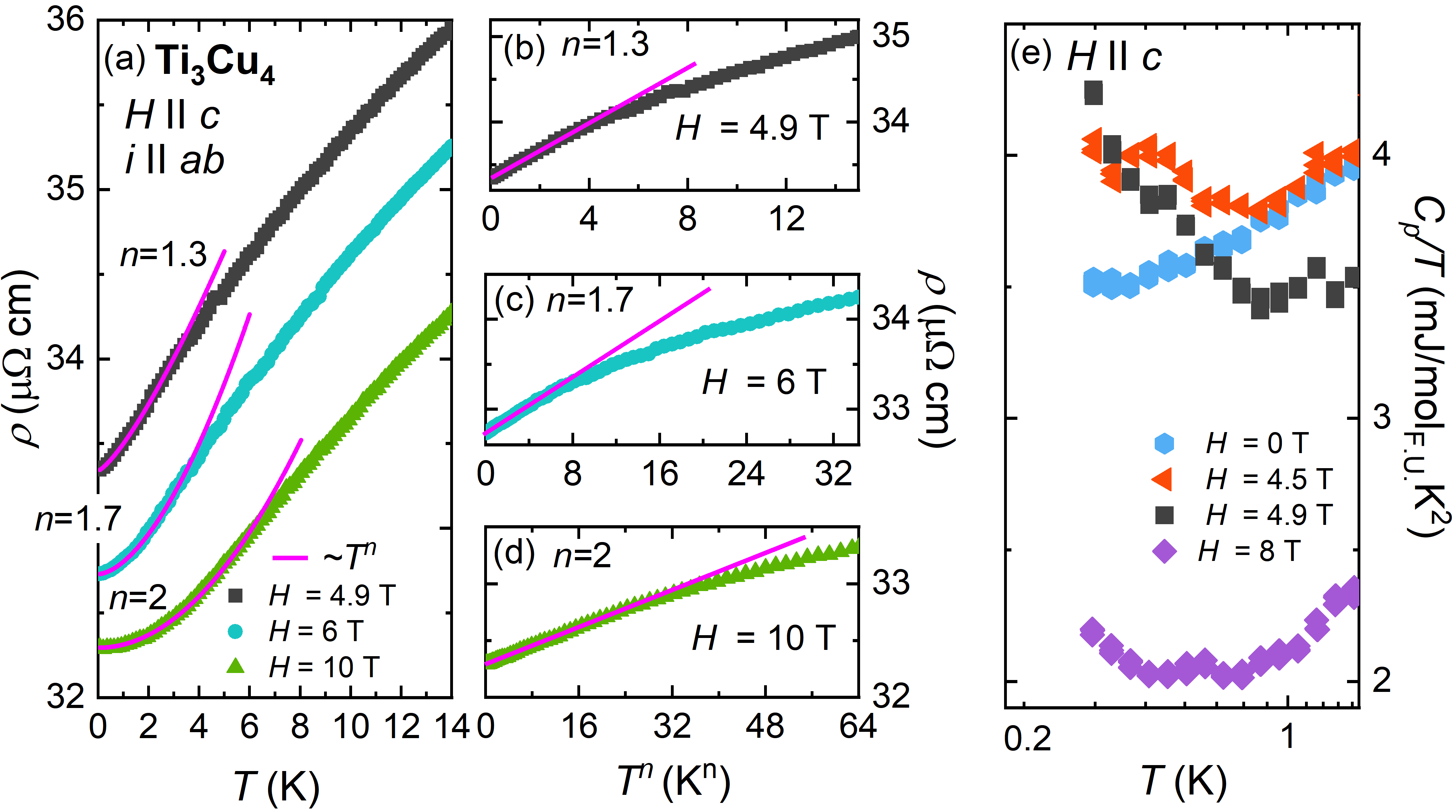}
  \caption{\textbf{Evidence for non-Fermi liquid behavior in Ti$_3$Cu$_4$.} (a) Temperature dependent resistivity measured from 50~mK$~<~T~<$~14~K for fields $H~\geq~H_c$ (closed symbols). Solid lines are fits to the lowest temperature data to the equation $\rho~=~\rho_0~+~AT^n$, where $n~=$~1.3 (NFL), 1.7 (intermediate) and 2 (FL), for $H~=~$4.9 T, 6 T, and 10 T, respectively. $\rho(T)$ data (symbols) are plotted against $T^n$ for (b) $n~=~$1.3 and $H~=~4.9$~T, (c) $n~=~$1.7 and $H~=~6$~T, and (d) $n~=~$1.7 and $H~=~10$~T. The pink lines are fits that show a single power corresponding to $n$, describe the resistivity for well over a decade in $T$ for all fields. (e) Heat capacity scaled by temperature $C_p/T$ on a semi-log $T$ scale from 0.3~K$~<~T~<$~1.5~K for fields 0~T$~<~H~<$~8~T.  A NFL divergence is seen near $H_c\sim 4.9$~T. See text for more details. } 
  \label{NFL}
\end{figure*}

Now turning to the electrical transport, we note that QCPs are often (albeit not always) accompanied by non-Fermi liquid (NFL) behavior, with a NFL-FL cross-over convergent at the QCP. Signatures of the NFL behavior are revealed by the resistivity analysis $\rho = \rho_0 + A_nT^{n}$, where the $T-H$ dependence of the exponent $n$ is represented by the contour plot in Fig.~\ref{fig:F4} for $H\parallel c$ for $T\geq~1.8$ K. A subset of the $\rho(T)$ data and fits to $\rho = \rho_0 + A_nT^{n}$ can be found in Supplementary Materials Fig. \ref{fig:Res}. At high temperatures in the paramagnetic state, $\rho(T)$ varies sub-linearly with temperature, \textit{i.e.}, $n < 1$. In other itinerant systems, similar behavior has been attributed to the conduction electrons being scattered by spin fluctuations of the $d$-band electrons \cite{ueda1977electrical,cao2004violation}. Just below  $T_N$, $n$ $\rightarrow$ 1 and $n\rightarrow$ 2 at the lowest measured $T$, for $H$ $\rightarrow$ 0. Above the QCP ($H>H_C$) as the temperature is lowered (Fig. \ref{fig:F4} and \ref{fig:Res}), $n$ crosses over to $n\approx$ 1 at intermediate temperatures, and to 2 on further cooling, signaling a FL regime at the lowest temperatures. Resitivity measured down to $T ~=~50$ mK for $H~\geq~H_c$, shown in Fig.~\ref{NFL}a as $\rho(T)$ and in Fig.~\ref{NFL}b-d as $\rho$ vs. $T^n$, demonstrate unambiguously a crossover from NFL (n $<$ 2) to FL behavior (n = 2) as $H$ moves away from $H_c$. Closest to $H_c$ (Fig. \ref{NFL}b), the exponent $n~=~1.3$ is registered over nearly two decades in $T$ from 50 mK to $\sim$ 3 K. Similarly, NFL (n = 1.7, Fig. \ref{NFL}c) or FL (n = 2, Fig. \ref{NFL}d) behavior occurs over two decades in temperature as the field increases up to 10 T.

Beyond transport, thermodynamic measurements reinforce the   NFL behavior with the divergence of the low $T$ specific heat (Fig. \ref{NFL}e). For metals, at low $T$, the electronic contribution to the specific heat is expected to dominate and the temperature dependence varies as $T^m$, where $m =$ 1 for a FL, and $m > 1$ is often associated with NFL behavior due to quantum fluctuations \cite{stewart2001non,lohneysen2007fermi}. Fig.~\ref{NFL}e shows $C_p/T$ plotted on a semi-log scale for 0.3 K $\leq~T~\leq$ 1.5 K at various fields 0 $ \leq~H~\leq$ 8 T. For $H~=0$, the data plateau towards the lowest temperature $T~=$ 0.3 K, as expected for a FL. A power-law divergence ($m > 1$) develops at $H~=$ 4.5 T and persists beyond H$_c$ up to $H~=$ 8 T, with the steepest divergence close to $H_c$ (black squares). Fig.~\ref{CPlow} in the Supplementary Materials shows evidence for a Schottky anomaly at the lowest temperatures, as $C_p/T$ increases on cooling starting at higher $T$ as $H$ is increased. However, for temperatures beyond those where the Schottky contribution is largest (T $>$ 0.3 K, Fig. \ref{NFL}e), the specific heat does not follow the trend expected from Schottky anomaly (no increase in T as H increases). The divergence in C$_p/T$ is therefore ascribed to NFL behavior, consistent with the transport measurements.

\section{Discussion}
Ti$_3$Cu$_4$ is an itinerant antiferromagnet for which the ordering temperature can be suppressed towards $T~=~$0 with a modest field resulting in a field induced QCP. This is therefore not only a new itinerant magnet with no magnetic elements, one of very few known to date, but also the first known such compound with a field-induced QCP. Typically, itinerant antiferromagnetism is associated with a strongly nested Fermi surface, where the nesting wavevector dictates the magnetic wavevector. Such mechanism applies to the prototypical itinerant antiferromagnet or SDW system, elemental Cr \cite{fawcett1988spin}. While the calculated Fermi surface for Ti$_3$Cu$_4$ appears nested in the $ab$ plane (Fig. \ref{fig:theo_main_fig}a-c), the experimental propagation wavevector points in the out-of-plane direction (Fig. \ref{OP}). An added conundrum is that the Fermi level lies on a sharp van-Hove singularity in the density of states (Fig. \ref{fig:theo_main_fig}e), which is often associated with itinerant ferromagnetism. A similar scenario was found in TiAu \cite{svanidze2015}, and it was later established that a new mechanism of mirrored van-Hove singularities in the Fermi surface separated by the experimentally determined magnetic wavevector lie at the origin of the itinerant antiferromagnetism \cite{goh2016mechanism,goh2017competing}. Further efforts are required to elucidate the novel origin of the magnetism in Ti$_3$Cu$_4$. 

From a quantum criticality perspective, Ti$_3$Cu$_4$ represents as a system without 4f electrons and is therefore free of the complication of f-d electron hybridization in the quantum critical regime. Since Fermi surface instabilities lie at the heart of itinerant magnetism, it is intuitive to understand how the effects of pressure or chemical doping may alter the Fermi surface, and in turn, the resulting magnetism or quantum criticality. However, it is less clear what the role of magnetic field is in tuning magnetism towards a QCP. Compared to f-electron systems, d-electron systems, have much larger energy scales associated with the magnetism which is reflected in their ordering temperatures ($T_{ord}$): $T_{ord}~\sim$ 0.1 to 5 K in the former, and 10-100's of K in the latter. Ti$_3$Cu$_4$ ($T_N~$=~11.3~K)  is unique in that the energy scale is seemingly small (a magnetic field of $H_c$~=~4.87~T can completely suppress the magnetism)  compared to TiAu ($T_N~$=~26~K) \cite{svanidze2015}  or Cr ($T_N~$=~311~K) \cite{fawcett1988spin} where magnetic fields have little effect on the magnetic ordering temperature. Ti$_3$Cu$_4$ therefore provides a model platform to study the of role magnetic fields as a tuning parameter for itinerant magnetic quantum criticality. It will be specifically informative to compare and contrast future studies when either chemical doping or pressure are used as the non-thermal control parameter. For example, doping Cr with V suppresses the magnetic order ending in a QCP \cite{lee2004high,jaramillo2009breakdown}, while Re and Ru \cite{nishihara1986superconductivity, chatani2003competition, matthias1962superconductivity, nishihara1985itinerant} suppress the magnetism resulting in a superconducting state which may be unconventional \cite{ramazanoglu2018suppression}.

\section{Conclusion}
In conclusion, Ti$_3$Cu$_4$ is a new itinerant AFM with no magnetic elements with $T_N = 11.3$~K and $\mu_{ord}~=~0.17 \mu_B/F.U.$ The magnetic state is remarkably fragile for a transition metal magnetic system, and can be suppressed to $T~=~0$ with a small applied field $H_C$ = 4.87 T, resulting in a field-induced QCP. Measurements of the  magnetic Gr{\"u}neisen ratio provide strong evidence for the quantum criticality, reinforced by the accompanied  NFL-FL crossover revealed by the resistivity and heat capacity measurements. Ti$_3$Cu$_4$ can serve as  as a platform for comparison and potentially generalization of the quantum critical behavior over the entire spectrum of magnetic moments from local to itinerant. In future studies, it will be important to understand the effects of pressure, chemical substitution, and disorder in Ti$_3$Cu$_4$, all of which are currently underway.

\section{Methods}

\TiCu{} was grown using a self flux method with a starting composition of Ti$_{0.33}$Cu$_{0.67}$. The constituent elements were arc melted and placed in a Ta crucible and sealed under partial argon pressure in a quartz ampoule. After the initial heating to 950$^\circ$C over 6 hours, a first step of fast cooling to 935$^\circ$C was followed by slow cooling to 895$^\circ$C over 62 hours, where the crystals were separated from the excess flux by spinning in a centrifuge. 

Single-crystal X-ray diffraction data on a \TiCu{} crystal were collected at 100(2) K with the use of a Bruker APEX2 \cite{bruker20091} kappa diffractometer equipped with graphite-monochromized MoK$_\alpha$ radiation ($\lambda$ = 0.71073 Å). The data collection strategy was optimized with the use of the algorithm COSMO in the APEX2 package as a series of $\omega$ and $\phi$ scans. Scans of 0.5º at 6 s/frame were used. The detector to crystal distance was 40 mm. The collection of intensity data as well as cell refinement and data reduction were carried out with the use of the program APEX2. The structure of \TiCu  was initially solved and refined with the use of the SHELX-14 algorithms of the SHELXTL program package \cite{sheldrick2008short}.  Face indexed absorption, incident beam, and decay corrections of the substructure were performed with the use of the program SADABS \cite{sadabs2008department}.  The program STRUCTURE TIDY \cite{gelato1987structure}  in PLATON \cite{spek2014multipurpose} was used to standardize the atomic positions of the substructure. Furthermore, powder diffraction was done using a Bruker x- ray diffractometer with Cu $\rm{K}_\alpha$ radiation. Powder and single crystal x-ray diffraction confirm the reported crystal structure for \TiCu~\cite{schubert1964einige}, apart from signs of mechanical stresses and possible minute ($<$ 5$\%$) Ta inclusions (non-magnetic). However, these do not affect the results of the current study on the magnetic properties of \TiCu. The crystallographic file in CIF format for the refined structure has been deposited with the Cambridge Crystallographic Data Centre as CCDC 1968322. These data may be obtained free of charge by contacting CCDC at (https://www.ccdc.cam.ac.uk/).

DC magnetization measurements were performed in a Quantum Design (QD) magnetic property measurement system from \textit{T} = 1.8 K - 300 K. The same system was used with a helium 3 insert for measurements from \textit{T} = 0.5 K - 1.8 K. Magnetization measurements up to $\mu_0H$ =30 T were carried out with an extraction magnetometer in a capacitor-powered pulsed magnet at the NHMFL pulsed field facility. The ac electrical resistivity measurements were made in a QD physical properties measurement system (PPMS) with a standard four-point probe technique for temperatures 2–300 K and magnetic field from 0-14~T. Measurements downt to 50 mK were made in the same instrument equipped with a dilution refridgerator.

Quasi-adiabatic magnetocaloric  effect (MCE) measurements between $0.25~$K$~<~T~<~1~$K were carried out in a QD PPMS equiped with a dilution refrigerator using the heat capacity option to ensure a quasi-adiabatic environment. The thermometer of a heat capacity puck with no sample mounted was calibrated as a function of field and temperature at several fields ranging from $0~<~H~<~14~$T. From this procedure, a calibration map was extablished for the thermometer resistance $R$, temperature $T$, and magnetic field $H$. The sample was then mounted and cooled using the heat capacity option to ensure that the sample temperature was at equilibrium with the bath temperature. $H$ was then swept at a rate of 105 Oe/s between 3.5 and 5.5 T and $R$ of the heat capacity thermometer was measured. Using the calibration map, the measured $R$ was converted to temperature, from which the MCE values were derived.

Muon spin relaxation ($\mu$SR) measurements were performed on a mosaic of single crystals at the M20 surface muon channel at TRIUMF. The crystals were mounted  on a low background sample holder with aluminum backed Mylar tape with their crystallographic $c$-axis parallel to the incident muon beam. Measurements were performed in the LAMPF spectrometer between 2 and 20~K in both longitudinal field geometry and in a weak ($H = 30$~G) transverse field. In this experiment, the total initial asymmetry, $A_0$, was determined by fitting the asymmetry spectra at high temperatures, in the weakly relaxing paramagnetic regime, giving $A_0 = 0.220$. Here we present the normalized muon polarization, $P(t) = A(t)/A_0$. Measurements were collected with the muon spins parallel to their momentum, such that the muons are implanted with their spins pointing along the $c$-axis, and also in spin-rotated mode, such that the muons are implanted with their spins lying within the crystals' $ab$-plane. No significant anisotropy was detected. The muon decay asymmetry spectra were fitted with a least squares minimization protocol using the muSRfit software package.

Single crystal elastic neutron scattering measurements were performed on the $E_i = 14.5$~meV fixed-incident energy triple axis spectrometer HB-1A at the High Flux Isotope Reactor, Oak Ridge National Laboratory. This experiment was performed with standard collimation settings (40’-40’-40'-80'), and the energy resolution at the elastic line was $\sim$~1 meV (full-width half-maximum). Adhesive was used to attach a 70~mg single crystal of Ti$_3$Cu$_4$ onto an aluminum plate. Measurements were performed in both the $(hk0)$ and the $(h0l)$ scattering planes. The crystal was oriented prior to the experiment at the CG-1B neutron alignment station. Measurements were performed at temperatures between 5~K and 20~K using a closed-cycle refrigerator. The magnetic symmetry analysis was performed with SARAh~\cite{wills2000new} and Rietveld refinements were carried out using FullProf~\cite{rodriguez1993recent}.

We performed Density Functional Theory (DFT) based calculations using the full-potential WIEN2K \cite{wien2k} and pseudo-potential ABINIT \cite{abinit} packages, with the generalized gradient approximation (GGA) used to account for the exchange-correlation interactions \cite{pbe}. The band structure, density of states and Fermi surfaces were computed with the full-potential WIEN2K code, whereas ABINIT was used to perform large supercell calculations to accommodate various spin-density wave (SDW) orders. We ensured that both the codes produced similar results at the level of the primitive unit cell. The polyhedron integration method was used to calculate the electronic density of states (DOS).

\section{Data Availability}
The data that support the findings of this study are available from the corresponding authors upon reasonable request.

\begin{acknowledgments}
\section{Acknowledgements}
We are grateful to Bassam Hitti and Gerald Morris for their assistance with the muon spin relaxation measurements at TRIUMF. We are also grateful to Anand B. Puthirath for help with some characterization, as well as Warren Pickett and Jeff Lynn for useful conversations.  We thank Ian Fisher and Pierre Massat for fruitful discussions on MCE measurements. JMM was supported by the National Science Foundation Graduate Research Fellowship under Grant DGE 1842494. CLH, SL and EM acknowledge support from NSF DMR 1903741. CLH is also supported by the Ministry of Science and Technology (MOST) in Taiwan under grant no. MOST 109-2112-M-006-026-MY3 and 110-2124-M-006-009. AMH, JB, YC, and GML were supported by the Natural Sciences and Engineering Research Council of Canada.
VL and AHN were supported by the Robert A. Welch Foundation grant C-1818.  AHN was also supported by the National Science Foundation grant no. DMR-1917511 and would like to thank for the hospitality of the Kavli Institute for Theoretical Physics, supported in part by the National Science Foundation under Grant No. NSF PHY-1748958. A portion of this work was performed at the National High Magnetic  Field Laboratory, which is supported by the National Science Foundation  Cooperative Agreement No. DMR-1644779, the State of Florida and the United States Department of Energy. Use was made of the Integrated Molecular Structure Education and Research Center X-ray Facility at Northwestern University, which has received support from the Soft and Hybrid Nanotechnology Experimental Resource (NSF Grant ECCS-1542205), the State of Illinois, and the International Institute for Nanotechnology. At Argonne, this work was supported by the US Department of Energy, Office of science, Basic Energy Sciences, Materials Sciences and Engineering Division (structural analysis). A portion of this research used resources at the High Flux Isotope Reactor, a DOE Office of Science User Facility operated by the Oak Ridge National Laboratory.

\end{acknowledgments}

\section{Author contributions}
E.M. designed the study. J.M.M. and K.B. grew the crystals. J.M.M. performed magnetization, transport, heat capacity, and magneto-caloric effect measurements. E.M., J.M.M., A.M.H., C.L.H. and S.L. performed the analysis and wrote the manuscript with contributions from all authors. V.L. and A.H.V. performed DFT calculations and analysis. A.M.H., J.B., Y. C. and G.M.L. performed muon spin relaxation measurements and data analysis. A.A.A, L.L.K., and A.M.H. performed elastic neutron scattering measurements and analysis. C.D.M and M.G.K were responsible for the structural characterization. F.W. measured the high field magnetization data.

\section{Competing financial interests}
The authors declare no competing financial interests.

\bibliography{Ti3Cu4}
\cleardoublepage

\pagebreak

\section*{Supplementary Materials}

\beginsupplement

\begin{figure}[H]
 \includegraphics[width=\columnwidth]{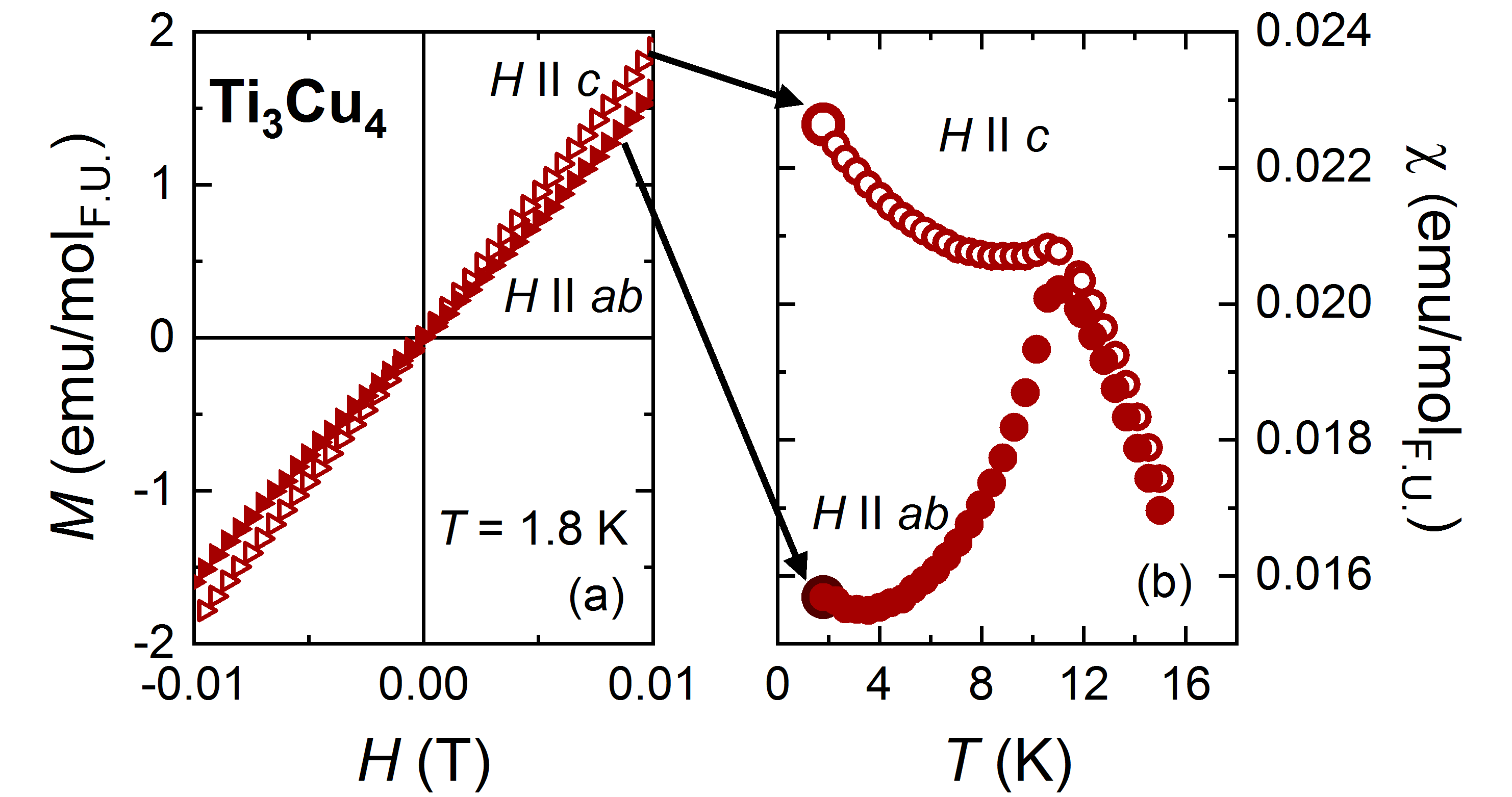}
  \caption{(a) Example of two magnetization isotherms used to determine $\chi$ at \textit{T}= = 1.8 K with $H\parallel c$ (open triangles) and $H\parallel ab$ (full triangles). (b) $\chi(T)$ determined from the slope of the low field $M(H)$ curves.  } 
  \label{fig:SF1}
\end{figure}

In order to investigate the magnetic anisotropy in \TiCu{} the zero field limit of the magnetic susceptibility is measured as $\chi = \displaystyle{\lim_{H \to 0}} \text{d}M/\text{d}H$. $M(H)$ isotherms were measured from $\mu_0H~=$ ~-0.01 T to 0.01 T at temperatures ranging from 1.8~K to 15.3~K. Two such isotherms are shown in Fig. \ref{fig:SF1}a for T =  1.8 K, with  $H || c$  (open triangles) and $H || ab$ (full triangles).  $\chi(T)$ is extracted from the slope of each isotherm and plotted vs. temperature in Fig. \ref{fig:SF1}b where the open circles correspond to the  field parallel to the $c$ axis, and full circles for field parallel to the $ab$ plane.

\begin{figure} [H]
 \includegraphics[width=\columnwidth]{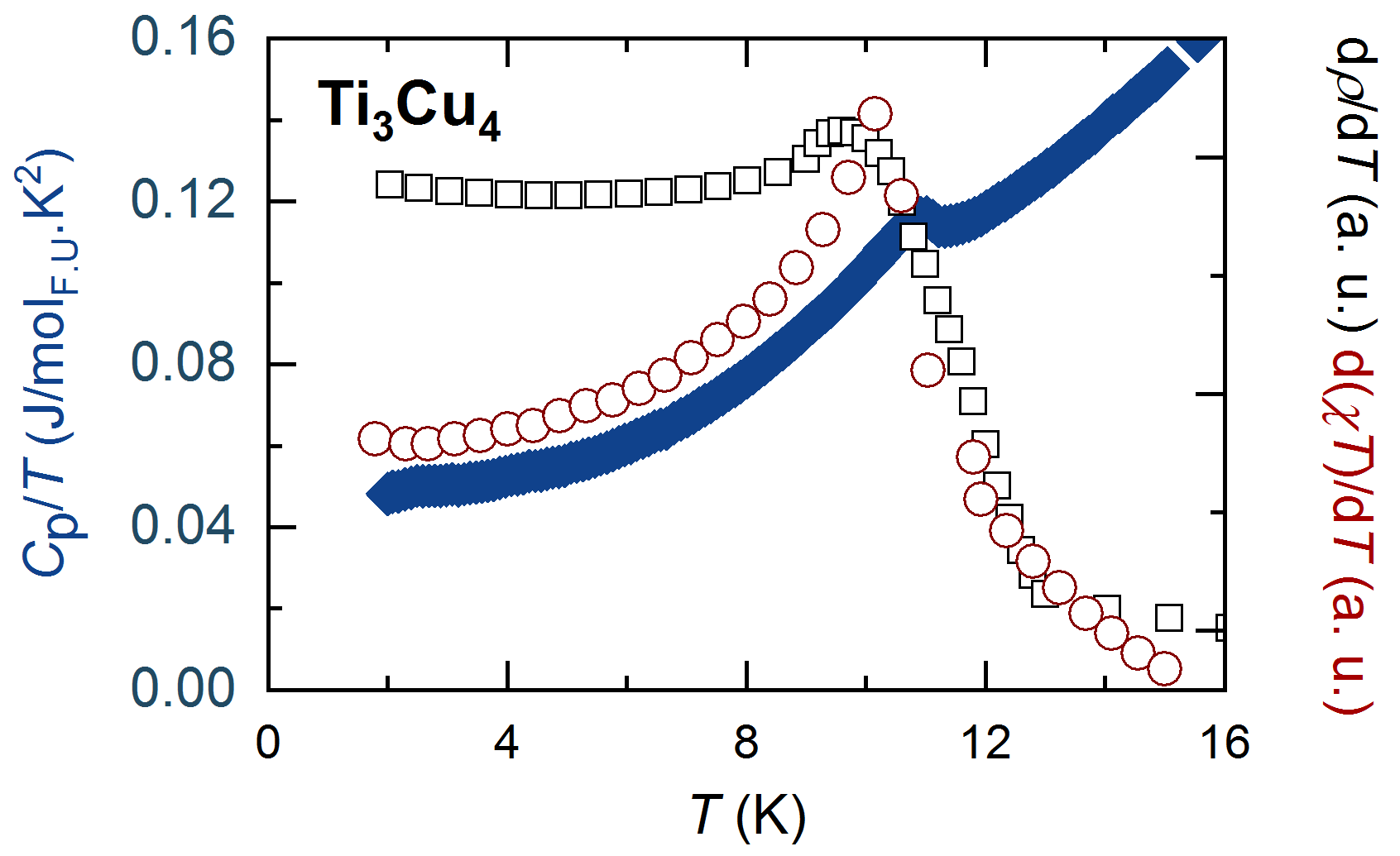}
  \caption{$T_N$ is extracted from the peaks in $C_p/T$ (left axis, blue diamonds), $\text{d}(\chi T)/\text{d}T$ (right axis, red circles), and $\text{d}\rho/\text{d}T$ (right axis, black squares). This panel presents data shown in  Fig. \ref{fig:F1}d-f.} 
  \label{fig:SF2}
\end{figure}

\begin{figure}
 \includegraphics[width=\columnwidth]{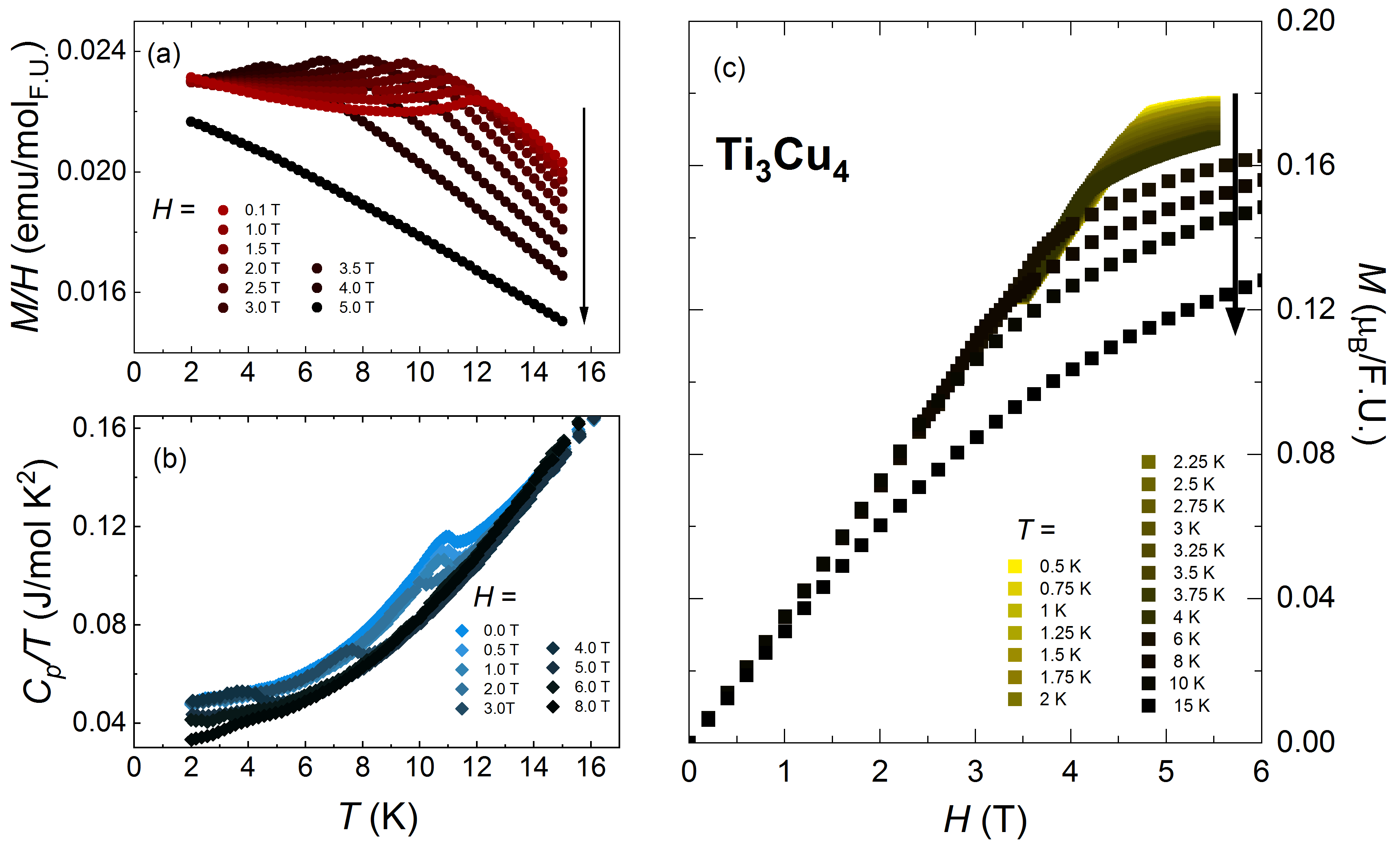}
  \caption{(a) Magnetic susceptibility (red circles) and (b) heat capacity scaled by temperature (blue diamonds) measured as a function of increasing field (red/blue to black). With increasing field,  $T_N$ is continually suppressed. (c) Magnetic isotherms measured between $T$=0.5-15 K (yellow to black squares). The metamagnetic transition continually increases in field as temperature is decreased. Magnetization measurements were performed with $H \parallel ab$, while heat capacity was measured with  $H \parallel c$.}
  \label{fig:SF3}
\end{figure}

\begin{figure}
 \includegraphics[width=\columnwidth]{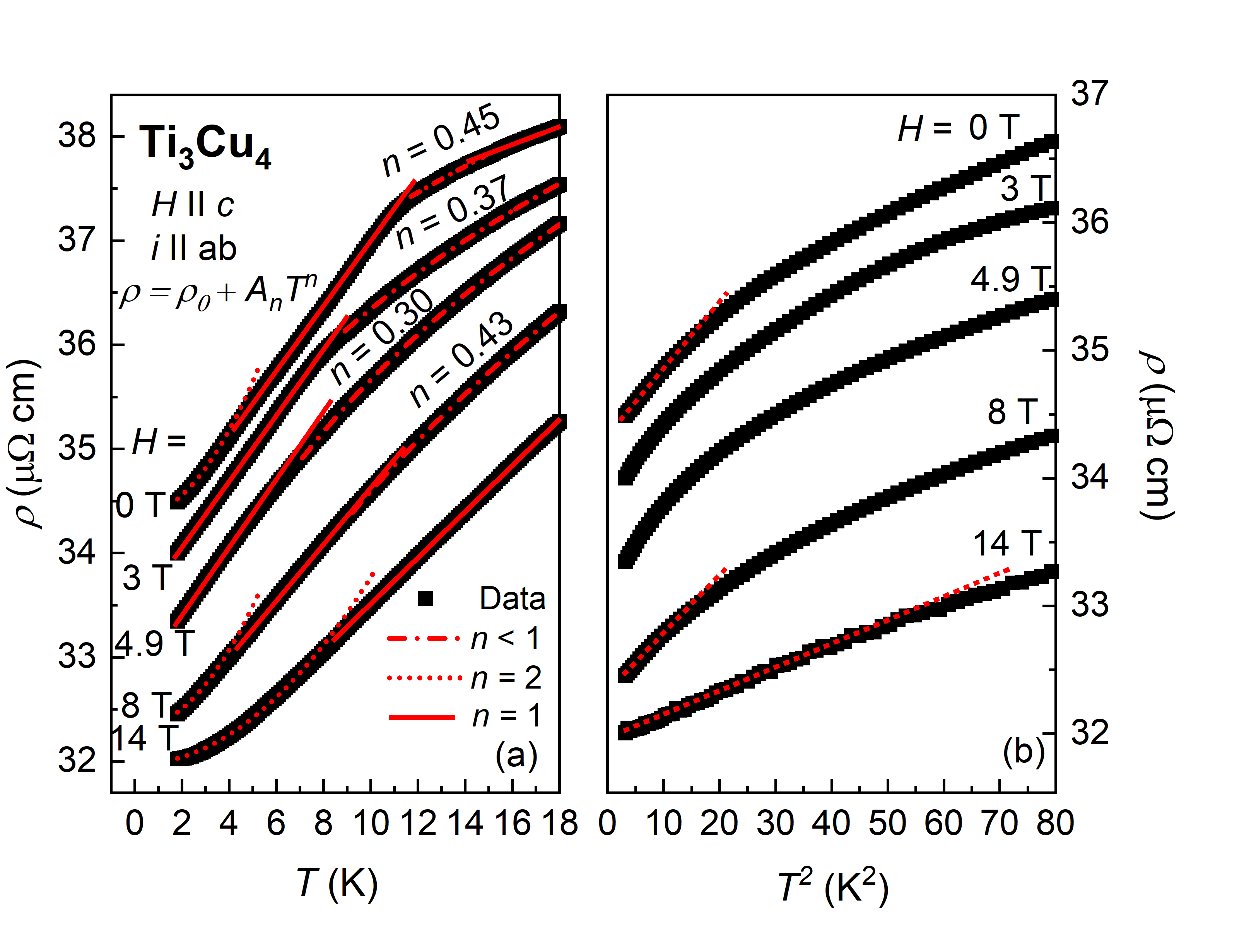}
  \caption{(a) Temperature dependent resitivity measurements with current \textit{i} $\parallel$ \textit{ab} and magnetic field \textit{H} $\parallel$ \textit{c} measured at \textit{H} = 0, 3, 4.9, 8, and 14 T plotted on a linear temperature scale. Red dotted curves correspond to the fitted exponent of $n =2$ when the resistivity is fit to the equation $\rho = \rho_0 +A_nT^n$. Red solid curves correspond to the fits where $n =1,$ and the dot-dashed lines correspond to fits where the extracted exponent $n < 1.$ (b) The lowest temperature data from (a) plotted against temperature squared, $T^2$.}
  \label{fig:Res}
\end{figure}

The AFM ordering temperature $T_N$ is extracted from the $C_p/T$ peak in Fig. \ref{fig:SF2} (left axis, blue diamonds), and the peaks in $\text{d}(\chi T/)\text{d}T$ (right axis, open circles), and $\text{d}\rho/\text{d}T$ (right axis, open squares). The phase diagram Fig. \ref{fig:F4} is constructed using the $M(T)/H$, $C_p/T$, and $M(H)$ manifolds shown in  Fig. \ref{fig:SF3}, while example curves of the $\rho(T)$ data are displayed in Fig. \ref{fig:Res}. It should be noted that for $H>0$, $T_N$ extracted from the magnetic susceptibility is determined by the derivative $\text{d}(MT)/\text{d}T$, from temperature sweeps of magnetization $M$ measured at constant field $H$. Interestingly, while the $M(H)$ data (Fig. \ref{fig:F2} and \ref{fig:SF2}) are consistent with a spin polarized state, a broad shoulder appears in the in $C_p/T$ at low $T$ most clearly seen in the $H = $ 8 T data in Fig. \ref{fig:F2}b. It is possible that the field-polarized state may have its own internal degrees of freedom causing the broad hump \cite{Reviewer1}.

\begin{figure}[tbp]
\linespread{1}
\par
\includegraphics[width=\columnwidth]{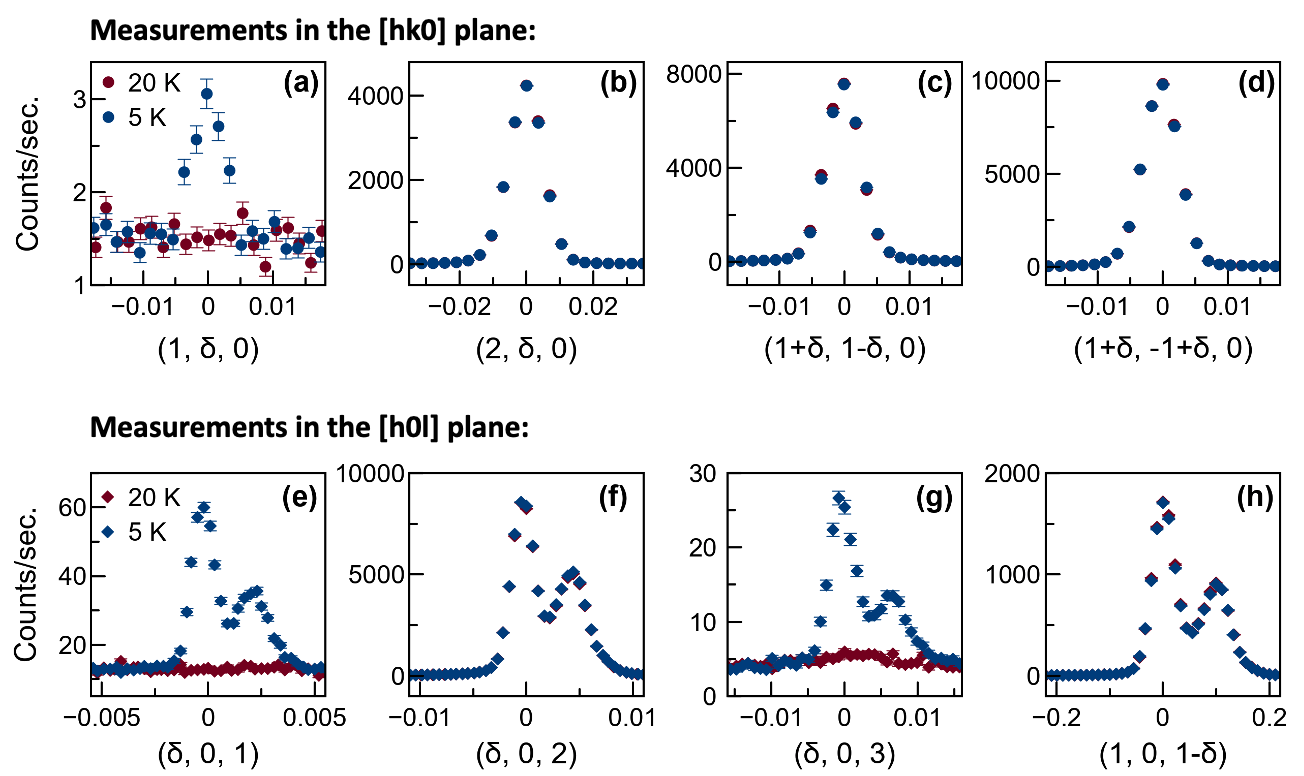}
\par
\caption{Rocking curve measurements performed in two different sample geometries. The (a) (100), (b) (200), (c) (110), and (d) (1$\bar{1}$0) Bragg positions were measured with the crystal aligned in the ($hk0$) plane at $T = 5$~K and 20~K. The (e) (001), (f) (002), (g) (003), and (h) (101) Bragg positions were measured with the crystal aligned in the ($h0l$) plane. Magnetic Bragg peaks form on the (100), (001), and (003) positions upon cooling through the $T_N = 11.3$~K transition. Note that the maximum divergence in the orthogonal direction for the rocking curves was of order 0.02\%.}
\label{BraggPeaks}
\end{figure}

\begin{figure}[h]
 \includegraphics[width=\columnwidth]{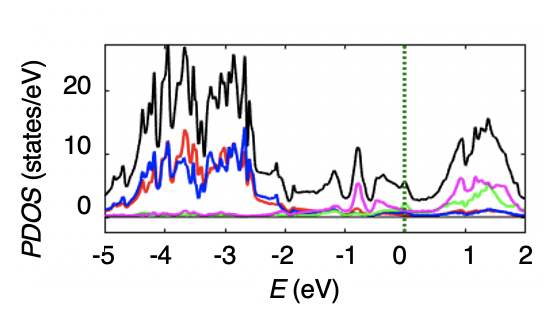}
 \caption{DOS plotted over a wide energy range. The total DOS is shown in black, while the colors show the partial DOS from various atoms. The Cu states (red and blue) lie well below the Fermi level, while the Ti states (green and pink) lie around the Fermi level.}
 \label{fig:Supp-dos}
 \end{figure}

The DOS at the Fermi level is dominated by orbitals of Ti, whereas the Cu states lie well below the Fermi level, as shown in Fig.~\ref{fig:Supp-dos}. In order to further determine the orbital composition of the DOS near the Fermi level, we have performed DOS calculations projected onto various orbitals (see Fig.~\ref{fig:theo_main_fig}e,f in the main text), with the conclusion that the largest contribution to the DOS at the chemical potential originated from Ti2 $d_{x^2-y^2}$ orbitals. These are the orbitals that play the decisive role in the magnetic ordering in this material.

\begin{figure}[b]
\vspace{5mm}
 \includegraphics[width=\columnwidth]{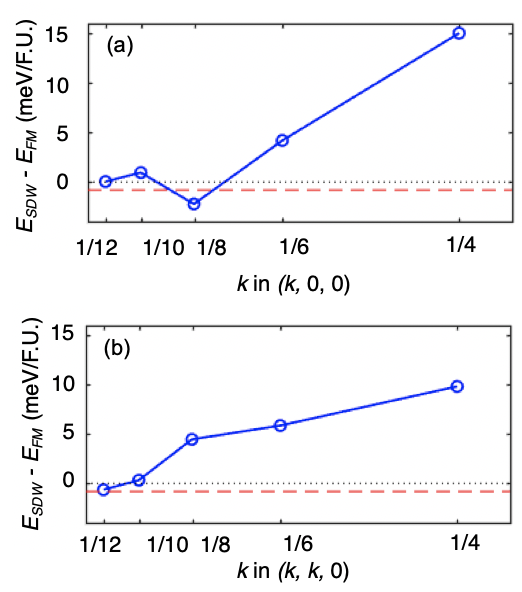}
  \caption{Energy dependence of the SDW on ordering wave-vector, \textbf{q} shown for (a) \textbf{k} $\parallel (1~0~0)$ and (b) \textbf{k} $\parallel (1~1~0)$. The dashed red line at -0.83 meV/f.u. indicates the energy of the experimentally realized antiferromagnetic state with ordering wavevector \textbf{k} = $(0~0~1)$. }
 \label{fig:Supp-sdw}
 \end{figure}

In order to determine the nature of magnetism in this material, we have performed a series of spin-density wave (SDW) calculations with different commensurate ordering wave-vectors parallel to $(1~0~0)$ and $(1~1~0)$ direction. 
The trial SDW states were implemented using the non-collinear ABINIT code, within an  enlarged supercell consistent with the pitch of a target SDW spiral state. The largest unit cell had dimension $12\times 12\times 1$, corresponding to the wavevector $\mathbf{k}=(1/12~1/12~0)$.
The comparison of the total energies of various competing SDW states is shown in Fig.~\ref{fig:Supp-sdw}, from which it follows that the lowest energy state is realized for the wavevector $\mathbf{k}=(1/8~0~0)$, corresponding to the wavelength $\lambda=8a$ of the magnetic spiral order. Interestingly, the experimentally determined ordered state with the wavevector $\mathbf{k}=(0~0~1)$ lies only marginally higher in energy, about 1 meV/f.u. (red dashed line in Fig.~\ref{fig:Supp-sdw}). This energy difference is however within the error bars of the DFT calculations and is not conclusive. What is evident from this analysis is that several candidate SDW states, including the experimentally observed one, are predicted to lie very close in energy. The approximations inherent in the DFT treatment of exchange and correlations do not allow us to predictably deduce which of these competing states is realized, and we instead rely on the neutron diffraction study (see main text) to deduce the ordered state with wavevector $\mathbf{k}=(0~0~1)$.

 \begin{figure}[t]
 \includegraphics[width=\columnwidth]{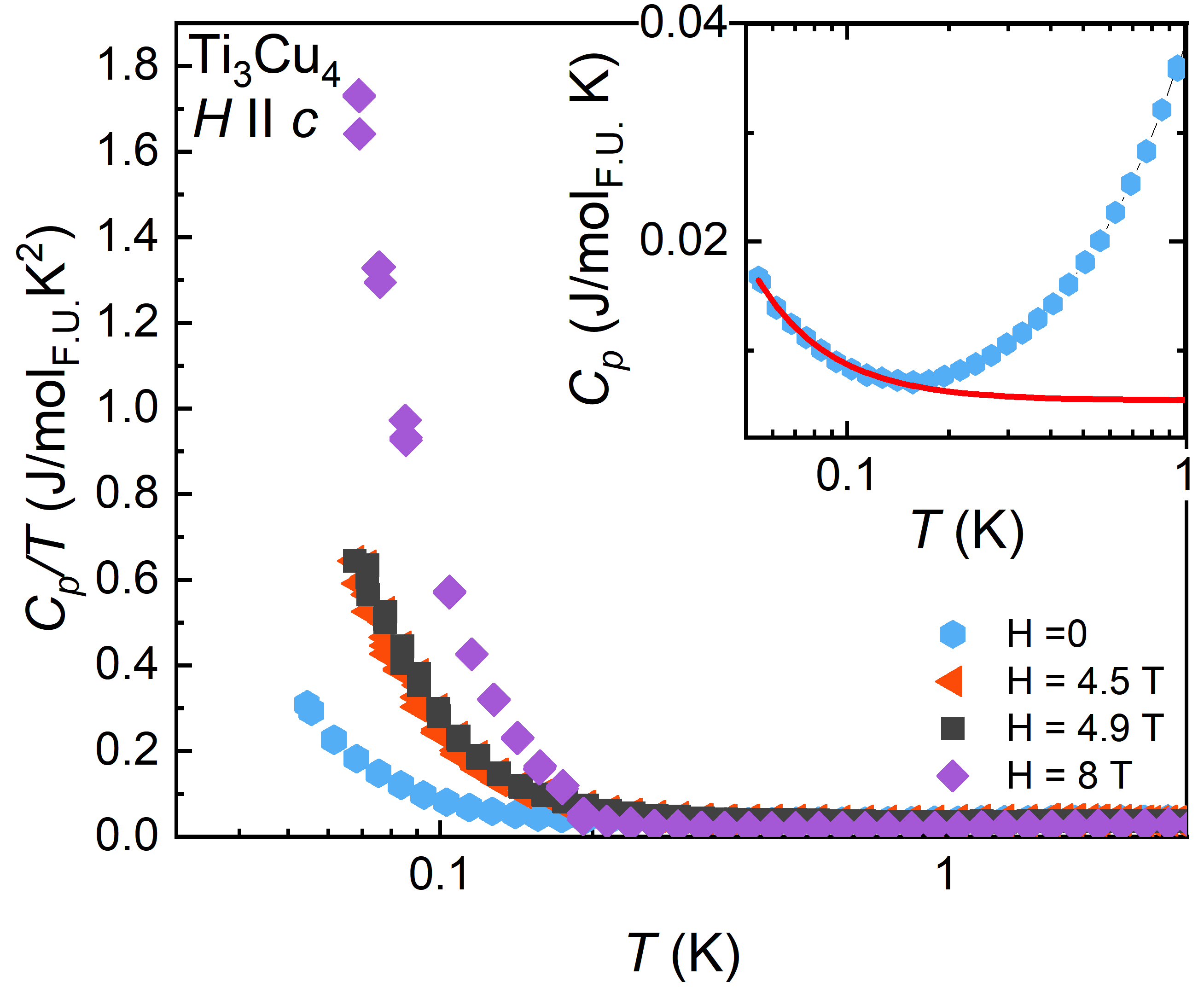}
  \caption{$C_p/T$ on a semi-log $T$ scale measured for 50 mK $\leq~T~\leq$ 3 K, showing an upturn on cooling attributed to nuclear Schottky. As expected, the onset of the Schottky anomaly increases in $T$ as $H$ is increased. Inset: fit (red line) to the expected Schottky contribution $C_{\rm {Schottky}} \propto \left({\frac {\alpha }{T}}\right)^{2}{\frac {e^{\alpha /T}}{[1+e^{\alpha /T}]^{2}}}$ for H = 0, where $\alpha$ is an $H$-dependent constant.} 
 
  \label{CPlow}
\end{figure}

\end{document}